\documentstyle[aaspp4]{article}

\received{}
\accepted{}
\journalid{VOL}{JOURNAL DATE}
\articleid{START PAGE}{END PAGE}
\paperid{MANUSCRIPT ID}

\slugcomment{}
\lefthead{Yoshida, Lee}
\righthead{}

\begin{document}

%%%%%%%%%%%%%%%%%%%%%%%%%%%%%%%%%%%%%%%%%%%%%%%%%%%%%%%%%%%%%%
%% O. FRONT MATTER                                          %%
%%%%%%%%%%%%%%%%%%%%%%%%%%%%%%%%%%%%%%%%%%%%%%%%%%%%%%%%%%%%%%

\title{Rotational modes of non-isentropic stars and the gravitational radiation 
  driven instability }
\author{Shijun Yoshida and Umin Lee}
\affil{Astronomical Institute, Graduate School of Science, 
Tohoku University, Sendai 980-8578, 
Japan \\ yoshida@astr.tohoku.ac.jp, lee@astr.tohoku.ac.jp}

\begin{abstract}
We investigate the properties of
$r$-mode and inertial mode of slowly rotating, non-isentropic, 
Newtonian stars, by taking account of the effects of 
the Coriolis force and the centrifugal force.
The Coriolis force is the dominant restoring force 
for both $r$-mode and inertial mode, which are also called rotational mode 
in this paper. 
For the velocity field produced by the oscillation modes
the $r$-mode has the dominant toroidal component over the 
spheroidal component, while the inertial mode has
the comparable toroidal and spheroidal components.
In non-isentropic stars the specific entropy of the fluid
depends on the radial distance from the center, and
the interior structure is in general divided into
two kinds of layers of fluid stratification that is
stable or unstable against convection.
Because of the non-isentropic structure, low frequency
oscillations of the star are affected by the buoyant force, which
has no effects on oscillations of isentropic stars.
In this paper we employ simple polytropic models with the polytropic index $n=1$
as the background neutron star models for the modal analysis.
For the non-isentropic models we consider only two cases, that is,
the models with the stable fluid stratification in the whole interior
and the models that are fully convective.
For simplicity we call these two kinds of models ``radiative'' and ``convective''
models in this paper.
For both cases, we assume the deviation of the models from isentropic 
structure is small.
Examining the dissipation timescales due to 
the gravitational radiation and 
several viscous processes for the polytropic neutron star models,
we find that the gravitational radiation driven instability of 
the nodeless $r$-modes associated with $l'=|m|$ remains strong 
even in the non-isentropic models,
where $l'$ and $m$ are the indices 
of the spherical harmonic function representing the angular dependence
of the eigenfunction.
Calculating the rotational modes of the radiative models 
as functions of the angular rotation frequency $\Omega$, we find that
the inertial modes are strongly modified 
by the buoyant force at small $\Omega$, where the buoyant force
as a dominant restoring force becomes
comparable with or stronger than the Coriolis force.
Because of this property we obtain the mode sequences 
in which the inertial modes at large $\Omega$
are identified as $g$-modes or the $r$-modes with $l'=|m|$ at small $\Omega$.
We also note that as $\Omega$ increases from $\Omega=0$
the retrograde $g$-modes become retrograde inertial modes,
which are unstable against the gravitational radiation reaction.
\end{abstract}


%%%%%%%%%%%%%%%%%%%%%%%%%%%%%%%%%%%%%%%%%%%%%%%%%%%%%%%%%%%%%%
%% I. INTRODUCTION                                          %%
%%%%%%%%%%%%%%%%%%%%%%%%%%%%%%%%%%%%%%%%%%%%%%%%%%%%%%%%%%%%%%

\section{Introduction}

Since the recent discovery of the gravitational radiation driven instability of
the $r$-modes by Andersson (1998) and Friedman \& Morsink (1998),
a large number of papers have been published on the properties of the 
rotational modes of rotating stars because of their possible importance 
in astrophysics
(e.g., Kojima 1998, Andersson, Kokkotas \& Schutz 1998, 
Kokkotas \& Stergioulas 1998, Lindblom \& Isper 1998, Beyer \& Kokkotas 
1999, Lindblom, Mendell \& Owen 1999, Kojima \& Hosonuma 1999, 
Lockitch \& Friedman 1999, Yoshida \& Lee 2000, Lockitch 1999, Yoshida 
et al 1999).
Here, we have used the term of ``rotational mode'' to refer to both $r$-mode and 
inertial mode of rotating stars.
As an important consequence of the instability, for example, Lindblom, Owen 
\& Morsink (1998) argued that due to the instability, 
the maximum angular rotation velocity of hot neutron stars is 
strongly restricted.  
Owen et al. (1998) also suggested that
the gravitational radiation emitted from hot young neutron stars due to
the $r$-mode instability would be one of the potential sources
for the gravitational wave detectors, e.g. LIGO.

The dominant restoring force for the rotational modes
is the Coriolis force and the characteristic frequency 
is comparable to the angular rotation frequency $\Omega$ of the star 
(e.g., Greenspan 1964, Pedlosky 1987). 
The $r$-modes are characterized by the properties that
the toroidal motion dominates the spheroidal motion in the velocity field, and that 
the oscillation frequency $\omega$ observed in the corotating frame of the star 
is given by $\omega=2m\Omega/(l'(l'+1))$ 
in the limit of $\Omega\rightarrow0$, 
where $m$ and $l'$ are the indices of the spherical harmonic function $Y_{l'}^m$ 
representing the toroidal component of the velocity field of the mode
(e.g., Papaloizou \& Pringle 1978 , Provost, Berthomieu \& Rocca 1981, 
Saio 1982). 
On the other hand, the inertial modes have the comparable toroidal and
spheroidal components of the velocity field and therefore they are not necessarily
represented by a single spherical harmonic function
(e.g, Lee, Strohmayer \& Van Horn 1992, Yoshida \& Lee 2000).  
Note that the inertial mode of this kind is also called ``rotation mode'' or 
``generalized $r$-mode'' (Lindblom \& Ipser 1998, Lockitch \& Friedman 1999).

Most of the recent studies on the $r$-mode instability
driven by the gravitational radiation reaction are restricted to 
the case of neutron stars with isentropic structure, since the barotropic
(one-parameter) equation of state, $p=p(\rho)$ is usually assumed for
both the interior structure and the small amplitude oscillation.
In isentropic stars
the specific entropy of the fluid in the interior is constant and
the buoyant force has no effects on the oscillation.
However, stars in general have non-isentropic interior structure, 
which may be divided into
two kinds of layers with fluid stratification 
stable or unstable against convection.
It is well know that the property of the oscillation modes of
non-isentropic stars is substantially different from that of isentropic stars,
since the buoyant force comes into play as a restoring force for 
low frequency oscillations like $g$-mode propagating in the layers with the stable 
stratification (see, e.g., Unno et al 1989).
This is also the case for rotational modes, which have low oscillation
frequencies comparable to the spin rate $\Omega$ of the star.
For example, we note that
the $r$-mode associated with $l'=|m|$ having the nodeless eigenfunction in the
radial direction is the only $r$-mode permitted in an isentropic star for a given $m$
(e.g, Lockitch \& Friedman 1999, Yoshida \& Lee 2000), 
but in non-isentropic stars with the stable fluid stratification all
the $r$-modes with $l'\ge |m|$ are possible for a given $m$, and there are
in principle an infinite number of $r$-modes for given $l'$ and $m$
(e.g., Provost et al 1981, Saio 1982, Lee \& Saio 1997). 
We therefore consider it important to investigate the modal properties of 
the rotational modes in non-isentropic stars, particularly in order to
answer the question whether the property of 
the $r$-mode instability for neutron star models with non-isentropic structure
remains the same as that for isentropic neutron star models.

In this paper, we study the $r$-mode and the inertial mode oscillations
of slowly rotating, non-isentropic, Newtonian stars, where
uniform rotation and adiabatic oscillation are assumed
to simplify the mathematical treatments. 
The effects of the rotational deformation of the equilibrium structure 
on the eigenfunction and eigenfrequency are included in the analysis, 
using the formulation by Yoshida \& Lee (2000) 
(see also Lee \& Saio 1986, Lee 1993).
We employ simple polytropic models 
with the polytropic index $n=1$
as the background neutron star models.
For the non-isentropic models we consider only two cases, that is,
the models that have stable fluid stratification in the whole interior
and those that are in convective equilibrium.
In this paper, following the conventions used 
in the field of stellar evolution and pulsation, 
we simply call these two kinds of models
``radiative'' and ``convective'' models, respectively, although
no energy transport in the interior is considered.
The plan of this paper is as follows. 
In \S 2, we briefly describe the basic equations employed in the linear modal analysis
of slowly rotating stars.
In \S 3, we show numerical results for the rotational modes of non-isentropic 
stars, and discuss the properties of the eigenfrequency and eigenfunction of the modes.
In \S 4, we examine the stability of simple neutron star models
against the $r$-modes, computing their growth or damping timescales due 
to the gravitational radiation reaction and some viscous damping processes.
\S 5 is for conclusions.

%%%%%%%%%%%%%%%%%%%%%%%%%%%%%%%%%%%%%%%%%%%%%%%%%%%%%%%%%%%%%%
%% II. Method of Solutions                                  %%
%%%%%%%%%%%%%%%%%%%%%%%%%%%%%%%%%%%%%%%%%%%%%%%%%%%%%%%%%%%%%%

\section{Method of Solutions}

The method of solutions used in this paper is exactly the same as that 
applied in Yoshida \& Lee (2000). Here, therefore, we briefly describe  
the basic equations and the method of calculation. 
The details are given in Yoshida \& Lee (2000).  

\subsection{Equilibrium Configurations}

We consider oscillations of uniformly rotating Newtonian stars. 
Thus the angular rotation frequency of the star, $\Omega$ is constant.
We assume the polytropic relation for the equilibrium structure of the model, 
and the relation between the pressure $p$ and the mass density $\rho$ is given by
\begin{equation}
p = K \, \rho^{1 + \frac{1}{n}} \, ,
\end{equation}
where $n$ and $K$ are the polytropic index and the structure constant determined by
giving the mass and the radius of the model, 
respectively. 
In this investigation, assuming slow rotation, we employ 
the Chandrasekhar-Milne expansion (see, e.g, Tassoul 1978) to obtain  
the equilibrium structure of rotating stars. 
In this technique the effects of the 
centrifugal force and the equilibrium deformation are treated 
as small perturbations to the non-rotating spherical symmetric structure 
of the star. 
The small expansion parameter due to rotation is chosen as 
$\bar{\Omega}=\Omega (R^3/G M)^{1/2}$, where $R$ and $M$ are 
the radius and the mass of the non-rotating star, respectively. 
The equilibrium configurations are constructed with accuracy up to the order of 
${\bar{\Omega}}^2$.  For simplicity, we consider a sequence of slowly 
rotating stars whose central density is the same as that of the 
non-rotating star.

\subsection{Perturbation Equations}

In the perturbation analysis, we introduce a parameter $a$ that is 
constant on the distorted effective potential surface of the star. In practice, 
the parameter $a$ is defined such that
\begin{equation}
\Psi (r,\theta) = \Psi_0 (a) \, , \label{ep-co}
\end{equation}
where $\Psi_0$ and $\Psi$ are the effective potentials for the non-rotating  
and slowly rotating models, respectively. Note that $\Psi_0$ is equal to  
the gravitational potential and $\Psi$ is the sum of the gravitational 
potential and the centrifugal potential. With this parameter $a$, 
the distorted equi-potential surface may be given by
\begin{equation}
r = a \lbrace 1 + \epsilon (a,\theta,\varphi) \rbrace \, ,
\label{def-a}
\end{equation}
where $\epsilon$ is the function of the order of 
${\bar{\Omega}}^2$, and is obtained by giving the functional forms
of $\Psi$ and $\Psi_0$ using the Chandrasekhar-Milne expansion (Yoshida \& Lee 2000). 
Here we have used spherical polar coordinates $(r,\theta,\varphi)$. 
However, we employ hereafter the parameter $a$ as the radial coordinate, 
in stead of the polar radial coordinate $r$.

The governing equations of nonradial oscillations of a rotating star are 
obtained by linearizing the basic equations used in fluid mechanics
(e.g., Unno et al 1989). 
In the following, we use $\delta$ and $\Delta$ to denote the Eulerian and 
the Lagrangian changes of physical quantities, respectively.
Since the equilibrium state is assumed to be stationary and axisymmetric, 
the perturbations may be
represented by a Fourier component proportional to 
$e^{i(\sigma t + m \varphi)}$, where $\sigma$ is the frequency observed in 
an inertial frame and $m$ is the azimuthal quantum number. 
The continuity equation may be linearized to be
\begin{equation}
\delta \rho = - \nabla_i (\rho \xi^i), \label{del_rho} \label{mass_con} 
\end{equation}
where $\xi^i$ is the Lagrangian displacement vector, and we have made use of
$\delta v^i  = i(\sigma + m \Omega) \xi^i \equiv i \omega \xi^i$
with $\omega$ being the oscillation frequency observed in the corotating 
frame of the star.  
Here $\nabla_i$ denotes the covariant derivative derived from the metric
$g_{ij}$. 

In this study, we consider only adiabatic oscillations. 
Thus, the perturbed equation of state is given by
\begin{equation}
\delta p    = \Gamma p \left( \frac{\delta \rho}{\rho} +
\xi^i A_i \right)  \, ,  \label{del_p} 
\end{equation}
where
$\Gamma$ is the adiabatic exponent defined as
\begin{equation}
\Gamma = \left( \frac{ \partial \ln p}{\partial \ln \rho} \right)_{ad} \, ,
\end{equation}
and $A_i$ is the generalized Schwarzschild discriminant\footnote{Since 
the Schwarzschild discriminant $A_i$ is proportional to $\nabla_i s$ with $s$
being the specific entropy, a model with $A_i=0$ 
throughout the interior is called isentropic model in this paper.
Using the same barotropic 
(one-parameter) equation of state, $p=p(\rho)$ for both
the equilibrium structure and the perturbation leads to isentropic models.} given by
\begin{equation}
A_i \equiv \frac{1}{\rho} \nabla_i \rho - \frac{1}{\Gamma p} \nabla_i p  \, ,
\label{def_A}
\end{equation}
and the positive and negative values of $A_i$ indicate the fluid stratification
unstable and stable against convection, respectively.

The linearized Euler's equation is
\begin{equation}
-\omega^2 \xi_i + \nabla_i 
\left( \frac{\delta p}{\rho} + \delta \Phi \right) + 
\left(\frac{\delta p}{\rho} \, g_{ij}+\xi_j \frac{1}{\rho} \nabla_i p
\right) \, A^j + 
2 i \omega \Omega \xi^j \nabla_j \varphi_i 
= 0 \, , \label{pert_Euler}
\end{equation}
where $\Phi$ denotes the gravitational potential, 
and $\varphi^i$ is the rotational Killing vector, with which the 3-velocity
of the equilibrium fluid of a rotating star is given as 
$v^i=\Omega\varphi^i$. The perturbed Poisson equation is
\begin{equation}
\nabla_i \nabla^i \delta \Phi = 4\pi G\delta \rho \, ,
\label{pert_Poisson}
\end{equation}
where $G$ is the gravitational constant.

Physically acceptable solutions of the linear differential equations are 
obtained by imposing boundary conditions at the inner and outer 
boundaries of the star.  The inner boundary conditions are the regularity 
condition of the perturbed quantities at the stellar center. 
The outer boundary conditions at the surface of the star are 
$\Delta p/\rho=0$
and the continuity of the perturbed gravitational potential at the surface to
the solution of $\nabla_i\nabla^i\delta\Phi=0$ that vanishes at the infinity.

In order to solve the system of partial differential equations given above, 
we employ the series expansion in terms of spherical harmonics to
represent the angular dependence of the perturbed quantities.
The Lagrangian displacement vector, $\xi^i$ is expanded in terms of the 
vector spherical harmonics as
\begin{equation} 
\xi^a = \sum_{l\geq\vert m \vert}^{\infty} a \, S_l(a)  
Y_l^m (\theta,\varphi) e^{i \sigma t} \, ,   \label{v_a}
\end{equation}
\begin{equation}
\xi^\theta = \sum_{l,l'\geq\vert m \vert}^{\infty} 
\left\{ H_l (a) {Y_l^m}_{,\theta}  + T_{l'} (a) \frac{1}{\sin \theta} \, 
{Y_{l'}^m}_{,\varphi} \right\} e^{i \sigma t} \, ,     \label{v_t}
\end{equation}
\begin{equation}
\xi^\varphi = \frac{1}{\sin^2 \theta} 
\sum_{l,l'\geq\vert m \vert}^{\infty} \left\{ H_l (a) {Y_l^m}_{,\varphi}
        - T_{l'} (a) {\sin \theta} \, {Y_{l'}^m}_{,\theta}
        \right\} e^{i \sigma t} \,   \label{v_p} 
\end{equation}
(Regge \& Wheeler 1957, see also Thorne 1980). 
The perturbed scalar quantity such as $\delta \Phi$ is expressed as
\begin{equation} 
\delta \Phi  = \sum_{l\geq\vert m \vert}^{\infty} \delta \Phi_l (a) \, 
Y_l^m (\theta,\varphi) e^{i \sigma t} \, .  \label{d_phi}
\end{equation}
Here, the expansion coefficients $T_{l'}$ are called ``axial''  
(or ``toroidal'') and the other coefficients are called 
``polar'' (or ``spheroidal''). It is known that
these two set of components are decoupled when $\Omega=0$.

Substituting the perturbed quantities expanded in terms of the spherical 
harmonics into the linearized basic 
equations ($\ref{del_rho}$), ($\ref{del_p}$), ($\ref{pert_Euler}$) 
and ($\ref{pert_Poisson}$), we obtain an infinite system of coupled linear 
ordinary differential equations for the expansion coefficients. 
Notice that the system of differential equations employed in this paper
is arranged to give
eigenfunctions and eigenfrequencies consistent up to the order of $\bar\Omega^3$ 
for slow rotation.
The explicit expressions for the governing equations are given in Appendix of 
Yoshida \& Lee (2000).

Since the equilibrium state of a rotating star is invariant under 
the parity transformation defined by $\theta \to \pi - \theta$, 
the linear perturbations have definite parity for that transformation.  
In this paper, a set of modes whose scalar perturbations such as 
$\delta \Phi$ are symmetric with respect to the equator is called ``even'' 
modes, while the other set of modes whose scalar perturbations are antisymmetric 
with respect to the equator is called ``odd'' modes 
(see, e.g., Berthomieu et al. 1978).
For positive integers $j=1,~2,~3,~\cdots$, we have $l=|m|+2j-2$ and 
$l'=l+1$ for even modes, and $l=|m|+2j-1$ and $l'=l-1$ for odd modes, 
where the symbols $l$ and $l'$ have been used to denote the spheroidal 
and toroidal components of the displacement vector $\xi^i$, 
respectively (see equations (\ref{v_a})--(\ref{d_phi})). 
Note that we do not use the term of the even (the odd) parity mode to refer to
the polar (the axial) mode, although this terminology is used traditionally in 
relativistic perturbation theory (e.g., Regge \& Wheeler 1957).

Oscillations of rotating stars are also divided into prograde and retrograde modes.
When observed in the corotating frame of the star,
wave patterns of the prograde modes are propagating
in the same direction as rotation of the star, 
whereas those of the retrograde modes are
propagating in the opposite direction to the rotation.
For positive values of $m$, the prograde and retrograde modes have negative and positive
$\omega$, respectively.

For numerical calculations, we truncate the infinite set of linear 
ordinary differential equations to obtain a finite set by discarding all 
the expansion coefficients associated with $l$ higher than $l_{max}$. 
This truncation determines the number of the expansion coefficients 
kept in the spherical harmonic expansion of each perturbed quantity. 
We denote this number as $k_{max}$. Our basic equation becomes a 
system of $4\times k_{max}$-th order linear ordinary differential equations, which,
together with the boundary conditions, are numerically solved as an 
eigenvalue problem of $\omega\equiv \sigma+m\Omega$ by using a Henyey type 
relaxation method (e.g., Unno et al . 1989, see also Press et al. 1992).

We cannot determine a priori $k_{max}$,
the number of the expansion coefficients kept in the eigenfunctions. 
In this paper, we start with an appropriate trial 
value of $k_{\mbox{\tiny max}}$ to compute an oscillation mode 
at a given value of $\bar\Omega$. 
Then, we increase the 
value of $k_{\mbox{\tiny max}}$, until the eigenfrequency converges within 
an appropriate numerical error, where the mode identification is done
by examining the form and the number of nodes of the dominant expansion coefficients.
We consider that the mode thus obtained
with the sufficient number of the expansion coefficients
is the correct oscillation mode we want for the rotating model at $\bar\Omega$.

In order to check our numerical code, we calculate eigenfrequencies of 
the $r$-modes of the $n=3$ and $\Gamma=5/3$ polytropic model 
at $\bar{\Omega}=0.1$. In Table 1, our 
results are tabulated together with those calculated by Saio (1982)  
by using a semi-analytical 
perturbative approach. As shown by Table 1, our results are in good 
agreement with Saio's results.

%%%%%%%%%%%%%%%%%%%%%%%%%%%%%%%%%%%%%%%%%%%%%%%%%%%%%%%%%%%%%%
%% III. Rotational Modes of Non-isentropic....              %%
%%%%%%%%%%%%%%%%%%%%%%%%%%%%%%%%%%%%%%%%%%%%%%%%%%%%%%%%%%%%%%

\section{Rotational Modes of Slowly Rotating Non-isentropic Stars}

Because it is believed that 
isentropic structure is a good approximation for young and hot neutron stars,
in this study we concentrate our attention to the case that 
the deviation of the non-isentropic models from isentropic structure 
is sufficiently small.
In this case, we may assume that the adiabatic exponent $\Gamma$ is given by
\begin{eqnarray} 
\frac{1}{\Gamma} = \frac{n}{n+1} + \gamma  \, ,
\label{def-gamma}
\end{eqnarray}
where $\gamma$ is a constant giving the deviation from the isentropic structure.
By substituting equation (\ref{def-gamma}) into equation (\ref{def_A}), 
we obtain an explicit form of $A_i$ as:
\begin{eqnarray}
A_i &=& - \, \gamma \, \nabla_i \ln p  \, .
\end{eqnarray}
In this study, we only consider oscillations of polytropic models with the index $n=1$.
Positive and negative values of $\gamma$ are respectively for convective and radiative
models.
Isentropic models are given by $\gamma=0$.
Since our formulation for stellar oscillations of rotating stars is 
accurate only up to the order of $\bar{\Omega}^3$, we shall apply our method 
to the oscillation modes in the range of $\bar\Omega\le 0.1$.

For convenience of later discussions, it may be useful to review
the properties of $r$-modes and inertial modes
of isentropic models with $\gamma=0$ (see, e.g., Yoshida \& Lee 2000). 
Because of 
the asymptotic behavior of the eigenfrequency of the rotational modes
in the limit of $\Omega\rightarrow 0$, it is 
convenient to introduce a non-dimensional frequency $\kappa$ as follows:
\begin{equation}
\frac{\omega}{\Omega} =  \kappa \, . \label{def-kappa}
\end{equation}
Since the rotational deformation is of the order of $\bar\Omega^2$,
the effects on the frequency of the rotational modes appear as the terms of 
the order of $\bar\Omega^3$. 
Therefore, we may expand the dimensionless 
eigenfrequency $\kappa$ of the rotational modes in powers of $\bar\Omega$ as follows: 
\begin{equation}
\kappa =  \kappa_0 + \kappa_2 \, \bar{\Omega}^2 
 + O(\bar{\Omega}^4)  \, , \label{def-kappa0}
\end{equation}
where
\begin{equation}
\kappa_0 = \frac{2 m}{l'(l'+1)} \, 
\label{r-fre}
\end{equation}
for the $r$-modes associated with the indices 
$l'$ and $m$ of vector harmonics with the axial 
parity (see, e.g. Papaloizou \& Pringle 1978).
For the inertial modes, there is no analytic expression for $\kappa_0$ in general.

As suggested by Provost et al. (1981) and Saio (1982),
the $r$-mode associated with $l'=|m|$
having the nodeless eigenfunction $T_{l'}(r)$ is the only $r$-mode permitted
for isentropic stars for a given $m$.
In the limit of $\Omega\rightarrow 0$, we have 
$\kappa\rightarrow\kappa_0=2m/l'(l'+1)$.
Among the expansion coefficients of the eigenfunctions 
the toroidal component $T_{l'=|m|}$ is dominating.
For the inertial modes,
the value of $\kappa$ in the limit of $\Omega\rightarrow 0$
is also finite but different from 
$\kappa_0 = {2 m}/{l'(l'+1)}$. 

The inertial modes are characterized by the two angular 
quantum numbers $l_0$ and $m$, where the value of $l_0$ is 
the maximum index $l$ of spherical harmonics associated with 
the dominant expansion coefficients 
of the eigenfunctions (see Lockitch \& Friedman 1999, Yoshida \& Lee 2000). 
Note that our definition of $l_0$ is not the same as that of Lockitch \& 
Friedman (1999), but the same as that of Lindblom \& Ipser (1998) and 
Yoshida \& Lee (2000).
Note also that odd (even) parity modes have odd (even) values
of $l_0-|m|$, and that
the odd parity mode having $l_0-|m|=1$ is the $r$-mode associated with $l'=|m|$.
The value of $\kappa_0$ for the inertial modes depends not only on the
indices $m$ and $l_0$, but also on the equilibrium structure 
of stars (e.g., Yoshida \& Lee 2000). 
For given values of $l_0$ and $m$, the number of different inertial modes is
equal to $l_0 - \vert m\vert(\ge 2)$. 
However, there seems to be no good quantum number to classify the different inertial
modes associated with the same $l_0$ and $m$.
We here introduce an ordering integer $n$ to use the labellings such as 
$i_n$ and $\omega_n$ for these inertial modes.
Since the frequencies of the inertial modes with the same $l_0$ and $m$
are found in a sequence given by, in the case of $m>0$,
\begin{equation}
\omega_{-(l_0 - \vert m\vert-1)/2} < \cdots < \omega_{-1} < 0 \leq  
\omega_{0} < \omega_{1} < \cdots < 
\omega_{(l_0 - \vert m\vert-1)/2}\, 
\end{equation}
for odd parity modes, and
\begin{equation}
\omega_{-(l_0 - \vert m\vert)/2} < \cdots < \omega_{-1} < 0 < 
\omega_{1} < \cdots  < 
\omega_{(l_0 - \vert m\vert)/2}\, 
\end{equation}
for even parity modes, we may define 
the ordering number $n$ as an integer that satisfies
the inequality $|l_0-|m|-1|/2\ge |n| \ge 0$ for odd parity inertial modes, and
$|l_0-|m||/2\ge |n| >0$ for even parity inertial modes
(see Lockitch \& Friedman 1998, Yoshida \& Lee 2000).
We apply this classification to the inertial modes of both isentropic and 
non-isentropic stars in the following discussions.

\subsection{$r$-modes}

For the radiative models, all the
$r$-modes associated with $l'\ge |m|$ are possible for a given $m$, 
and there are in principle 
a countable infinite number of $r$-modes for given $l'$ and $m$. 
The $r$-modes with the same $l'$ and $m$ may be classified by counting
the number of nodes
of the axial components $T_{l'}$ in the radial direction.
For them the values of $\kappa_2$ are different, but the values of $\kappa_0$
are the same and equal to $2m/l'(l'+1)$.
It is important to note that this classification 
of the $r$-modes is applicable only in the limit of $\bar\Omega\rightarrow 0$
and $\kappa\rightarrow \kappa_0=2m/l'(l'+1)$, 
since 
the components associated with the other values of $l'$ will come into play 
for large $\bar\Omega$
(see, e.g. Provost et al. 1981, Saio 1982).
Let us introduce an integer $k$ to denote
the number of nodes of the toroidal eigenfunction $T_{l'}$,
where the node at the center of the 
star is excluded from the count.
Then, the $r$-modes are 
completely specified by the three quantum numbers $m$, $l'$ and $k$. 
In this paper, for given $l'$ and $m$,
we employ the notation of $r_k$ to denote the $r$-mode whose 
toroidal eigenfunction $T_{l'}(r)$ has $k$ nodes in the radial direction.

In Figures 1 to 3, the dimensionless eigenvalues $\kappa$ are given versus 
$\bar{\Omega}$ for several 
low-radial-order $r$-modes with $m=1$, 2, and 3
for the $n=1$ polytropic model with $\gamma=-10^{-4}$, where
the $r$-modes with $l'= |m|$, $l'= |m| + 1$ and 
$l'= |m| + 2$ are given in each figure.
Modes with $l'= |m| + 1$ 
are even parity modes, and those with $l'=|m|$ and $l'=|m|+2$ are odd parity modes. 
In the figures, the eigenvalues $\kappa$ of the $i_0$-modes with $l_0=|m|+2 k+1$,
calculated at $\bar{\Omega}=0.1$ 
for the $n=1$ isentropic model, are designated by
the filled circles, where the integer $k$ corresponds to the number in the subscript 
of the notation $r_k$.
Figures 1 to 3 clearly show that
the qualitative behavior of the eigenfrequencies as functions of $\bar\Omega$
is not strongly dependent on the azimuthal quantum number $m$. 
From these figures we also understand that
the $r$-modes of the radiative model with $\gamma < 0$
are divided into three distinctive classes according to the asymptotic 
behavior of the eigenvalues at large $\bar\Omega$.
The first class of the $r$-modes consists of the nodeless $r_0$-modes 
associated with $l'=|m|$. 
The eigenvalue $\kappa$ of the $r$-modes of this class is well approximated by 
equations (17) and (18) at any values of $\bar\Omega$ examined in this paper.
Note that the $r_0$-mode with $m=l'=1$ is a 
peculiar mode in the sense that $\kappa$ is almost constant since
$\kappa_2=0$ and $\kappa=\kappa_0+O(\bar\Omega^4)$ (see Saio 1982, Greenspan 1964).  
The second class consists of the $r_k$-modes with 
$l'=\vert m \vert$ and $k>0$. 
Equations (\ref{def-kappa0}) and (\ref{r-fre}) can give good values for 
the eigenvalue $\kappa$ of the $r$-modes of this class
only when $\bar{\Omega}$ is sufficiently small.
As $\bar\Omega$ increases, $\kappa$ tends to
the $i_0$ inertial modes with $l_0=|m|+2 k+1$ of the isentropic model. 
Note that the $i_0$-modes are odd parity inertial modes 
(see equations (19) and (20)).
We therefore consider that at large $\bar\Omega$ 
the $r_k$-modes of this class become
identical with the $i_0$-modes with $l_0=|m|+2 k+1$, for which $\kappa$
is almost constant.
The third class of the $r$-modes consists of the $r_k$-modes with 
$l' > \vert m \vert$ and $k\ge0$.
The eigenvalue $\kappa$ of the $r$-modes of this class is well represented by
equations (17) and (18) only when $\bar\Omega$ is sufficiently small, and
it becomes small monotonically as $\bar\Omega$ increases.
The $r$-modes of this class may have no corresponding inertial modes
of the isentropic model.

As proved by Friedman \& Schutz (1978), oscillation modes whose frequency $\sigma$
observed in an inertial frame
satisfies the condition $\sigma(\sigma+m\Omega)=\sigma\omega<0$   
are unstable to the gravitational radiation reaction. 
The condition for the instability
may be rewritten as $0 < \kappa <  m$ for retrograde modes with $\omega>0$ and $m>0$.  
Figures 1 to 3 show that the $r$-mode with $l'\geq2$ 
satisfies the inequality $0 < \kappa < m$ for the range of $\bar\Omega$
examined in this paper.
It is also found that the $r_k$-modes with $l'=1$ 
and $k>0$ also satisfy the inequality $0 < \kappa < m$ for $\bar\Omega>0$.
The reason for the instability of the $r_k$-modes with $l'=1$ and $k>0$ may be
understood by noting that the 
polar components with the index $l=2$ can be responsible for the emission 
of gravitational radiation. 
We may conclude that the $r_0$-mode with $l'=\vert m\vert=1$ is 
the only $r$-mode that is stable against 
the gravitational radiation reaction.

In Figure 4, the eigenfrequencies $\kappa$ of three $r$-modes with $m=2$
are plotted as functions of $|\gamma|$, where
we have assumed $\bar\Omega=0.01$.  
The three modes are respectively the $r_0$-mode with $l'=m=2$, 
the $r_1$-mode with $l'=m=2$, and the $r_0$-mode with $l'=3$ and $m=2$.
These three $r$-modes belong to the first, second, and third classes of the $r$-modes
discussed in the previous paragraph.
The solid lines and the dotted 
lines are used to denote the cases of negative and positive $\gamma$, 
respectively.
Note that although we can obtain the three $r$-modes without any difficulty
for the radiative models with negative $\gamma$, 
we cannot always calculate the $r$-modes of the second and the third classes
for the convective models with positive $\gamma$.
Let us first discuss the case of the radiative models.
As shown by Figure 4, the eigenvalue $\kappa$ of the $r_0$-mode with $l'=|m|=2$ 
is practically independent of $|\gamma|$ and has values approximately equal to
$\kappa_0=2m/l'(l'+1)=2/3$.
However, $\kappa$ of the $r_1$-mode with $l'=|m|=2$ 
is dependent on $|\gamma|$.
At small $|\gamma|$, the $r_1$-mode has the eigenvalue close to $\kappa_0=0.517$, 
which corresponds to  
the inertial mode with $l_0-\vert m \vert = 3$ of the isentropic model, but 
it tends to $2m/l'(l'+1)=2/3$ as $|\gamma|$ increases.
The $r_0$-mode with $l'=3$ and $m=2$ also has the eigenfrequency $\kappa$
which depends on $|\gamma|$.
Although the value of $\kappa$ is close to $\kappa_0=2m/l'(l'+1)=1/3$ 
at large $|\gamma|$, it tends to zero as $|\gamma|\rightarrow 0$.
This behavior of $\kappa\rightarrow 0$ in the limit of $|\gamma|\rightarrow 0$
is consistent with the fact that the $r$-modes associated with 
$l'> \vert m\vert$ do not exist in isentropic stars, for which $\gamma=0$. 
Let us next discuss the case of the convective models with $\gamma>0$.
The eigenvalue $\kappa$ of the $r_0$-mode with $l'=|m|=2$ is only weakly dependent on
$\gamma$, as in the case of the radiative models.
Note that the curves for the $r_0$-mode with $l'=|m|=2$ for negative and positive $\gamma$
are almost indistinguishable in Figure 4.
The $r_1$-mode with $l'=|m|=2$ for the convective models behaves differently than 
that for the radiative models as $|\gamma|$ increases.
In fact, the value of $\kappa$ decreases
rapidly with increasing $\gamma$ and 
the $r_1$-mode cannot be found at large values of $\gamma>0$.
No $r_0$-mode with $l'=3$ and $m=2$ are found for the convective models.

Let us discuss the properties of the eigenfunctions of the $r$-modes 
for the radiative models.
Since the amplitude of the displacement vector
is much larger than that of the scalar perturbations, in the following discussions
we may show only the displacement vector of the modes.
In Figures 5 to 7, we show the expansion coefficients 
$S_l$, $H_l$ and ${\it i} T_{l'}$ as functions of the fractional radius $a/R$
for the $r$-modes with $m=2$, where we have assumed $\gamma=-10^{-4}$, and
the amplitude is normalized at the surface such that $iT_{l'}=1$. 
Here, $l=|m|$ and $l'=l+1$ for even modes, and $l=|m|+1$ and $l'=l-1$ for odd modes.
The modes given in Figures 5 to 7 are respectively
the $r_0$-mode with $l'=2$, the $r_1$-mode with $l'=2$, and the $r_0$-mode with $l'=3$. 
In each figure, the eigenfunctions for the cases of 
$\bar{\Omega}= 10^{-3}$ and $\bar{\Omega}= 10^{-1}$ 
are shown in panels $(a)$ and $(b)$, respectively. 
The panels $(a)$ of Figures 5 to 7 show that 
the toroidal component $iT_{l'}$ of the $r$-modes is dominating the other expansion 
coefficients when $\bar\Omega$ is small. 
However, as shown by the panels (b), although $iT_{l'}$
of the $r_0$-mode with $l'=2$ remains dominant even at 
$\bar{\Omega}= 0.1$, the toroidal components $iT_{l'}$
of the $r_1$-mode with $l'=2$ and the $r_0$-mode with $l'=3$
are not necessarily dominating any more at $\bar{\Omega}= 0.1$.
The properties of the eigenfunctions of the $r_1$-mode with $l'=m=2$ 
at large $\bar\Omega$ may be regarded as those of the inertial modes. 
It is interesting to note that
the displacement vector of the $r_0$-mode with $l'=3$ has large amplitude 
in the deep interior and almost vanishing amplitude near the stellar 
surface at large $\bar\Omega$. 
This suggests that because of the low frequencies $\omega$ in the corotating frame
the buoyant force is as important as the Coriolis force to determine the
mode character at large $\bar\Omega$.

\subsection{Inertial modes}

For the radiative models with the polytropic index $n=1$ and $\gamma = -10^{-4}$,
the eigenfrequencies $\bar{\omega}\equiv\omega/(GM/R^3)^{1/2}$ 
of low frequency oscillations
are shown as functions of $\bar\Omega$
for $m=1$, 2, and 3 in Figures 8 to 10,
where even and odd parity modes are displayed in panels (a) and (b), respectively.
The straight dashed lines drawn in the figures are given by
$\bar{\omega}=0$ and $\bar{\omega}= m \bar{\Omega}$ (see below).
In the figures, the inertial modes of the $n=1$ isentropic model 
computed at $\bar\Omega=0.1$ are designated by the filled circles.
We note that the oscillation modes in a mode sequence
are regarded as inertial modes when $\bar\Omega$ is large, and
as internal gravity ($g$-) modes when $\bar\Omega$ is small.
If we follow one of the continuous mode sequences from large $\bar\Omega$ 
to small $\bar\Omega$,
we may physically understand the behavior of the low frequency oscillation 
in the radiative models as a function of $\bar\Omega$.
At large $\bar\Omega$, the Coriolis force is dominating as the restoring force
and the oscillation mode is identified as an inertial mode.
As $\bar\Omega$ decreases, however, the buoyant force 
becomes dominating the Coriolis force
and the mode becomes identical with a $g$-mode 
in the limit of $\bar\Omega\rightarrow 0$.
Because of the effects of the buoyant force in the radiative model,
the frequency $\bar\omega$ of the inertial modes deviates from that
of the inertial modes of the isentropic model as $\bar\Omega$ decreases.

The modes within the two straight dashed lines depicted in Figures 8, 9 
and 10 are
unstable against the gravitational radiation driven instability 
in the sense that they satisfy the frequency condition
$\sigma(\sigma+m\Omega) < 0$ (Friedman \& Schutz 1978).
We can see from Figure 8 that the $i_k$-modes with $k\neq0$ and 
$m=1$ are stable against the instability for all $\bar{\Omega}$. 
Figures 9 and 10 show that as $\bar\Omega$ increases
all the retrograde $g$-modes with $|m|\geq 2$, which are stable 
against the instability when $\bar\Omega\sim0$, 
become retrograde inertial modes, which are unstable against the instability 
at large $\bar{\Omega}$.

There are several cases where we find it practically impossible to
obtain completely continuous mode sequences 
between the $g$-modes at small $\bar\Omega$ and
the inertial modes at large $\bar\Omega$.
In such sequences, there exist a domain of $\bar\Omega$, in which 
there appear a lot of modes having a large number of nodes
of the dominant expansion coefficients, and it is difficult to identify the
same oscillation modes as those computed in the previous steps in $\bar\Omega$.
In Figures 8 to 10, the domains of $\bar\Omega$ with this difficulty are indicated by
the dotted lines in the mode sequences.
Unfortunately, we cannot remove the difficulty by increasing $k_{max}$.
One of the reasons for the difficulty may be attributed to the method of
calculation we employ in this paper.
As discussed in Unno et al (1989), since we have truncated 
the infinite system of differential equations to obtain a finite system of equations,
the numerical procedure used here inevitably contains
a process of inverting matrices which are singular at some value of the ratio
$2\Omega/\omega\sim 1$, depending on the oscillation mode we are interested in.
In this paper, we assume that there always exist a continuous mode sequence
between the $g$-mode and inertial mode as a function of $\bar\Omega$, 
even if the mode sequence contains
a domain of $\bar\Omega$ in which the modes we are interested in cannot be
obtained numerically.

\subsection{Connection rules between the $g$-modes, $r$-modes and inertial modes}

In order to find 
the connection rule between the $g$-modes at small $\bar\Omega$ and 
the inertial modes at large $\bar\Omega$, we tabulate first the
frequencies of the $g$-modes at $\bar\Omega=0$ 
for the radiative model with the index $n=1$ and $\gamma=-10^{-4}$ in Table 2.
To obtain a clear understanding of the global correspondence between 
the $g$-, $r$-, and inertial modes, with the help of Table 2 and Figures 8 to 10, 
we give schematic diagrams
for the connection rules between the several low-radial-order modes as follows: 

\noindent
For odd parity modes we have
\begin{eqnarray}
\footnotesize
&\left[
\begin{array}{llll}
                                 &                                & &\cdots\\
                                 &                                &
i_{2\phm{-}}(l_0=\vert m\vert+5)&\cdots\\ 
                                 &i_{1\phm{-}}(l_0=\vert m\vert+3)&
i_{1\phm{-}}(l_0=\vert m\vert+5)&\cdots\\
i_{0\phm{-}}(l_0=\vert m\vert +1)&i_{0\phm{-}}(l_0=\vert m\vert+3)&
i_{0\phm{-}}(l_0=\vert m\vert+5)&\cdots\\ 
                                 &i_{-1}(l_0=\vert m\vert +3)     &
i_{-1}(l_0=\vert m\vert +5)     &\cdots\\
                                 &                                &
i_{-2}(l_0=\vert m\vert +5)     &\cdots\\
                                 &                                & &\cdots\\
\end{array} 
\right]& \label{ina-odd}\\
& & \nonumber\\
&\Updownarrow& \nonumber\\
& & \nonumber\\
&\left[
\begin{array}{llll}
 & & &\cdots\\
 & &g_{1\phm{-}}(l_{\phm{0}}=\vert m\vert+3)&\cdots\\ 
 &g_{1\phm{-}}(l_{\phm{0}}=\vert m\vert+1)&
g_{2\phm{-}}(l_{\phm{0}}=\vert m\vert+1)&\cdots\\
r_{0\phm{-}}(l'_{\phm{0}}=\vert m\vert)\phm{+1}& 
r_{1\phm{-}}(l'_{\phm{0}}=\vert m\vert)\phm{+3}&
r_{2\phm{-}}(l'_{\phm{0}}=\vert m\vert)\phm{+5}&\cdots\\
 &g_{-1}(l_{\phm{0}}=\vert m\vert+1)&
g_{-2}(l_{\phm{0}}=\vert m\vert+1)&\cdots\\
 & &g_{-1}(l_{\phm{0}}=\vert m\vert+3)&\cdots\\
 & & &\cdots\\
\end{array}
\right]& \ , \label{gra-odd}
\end{eqnarray}
and for even parity modes we have
\begin{eqnarray}
\footnotesize
&\left[
\begin{array}{llll}
 & & &\cdots\\
                              & &i_{3\phm{-}}(l_0=\vert m\vert+6)&\cdots\\ 
                                &i_{2\phm{-}}(l_0=\vert m\vert+4)&
i_{2\phm{-}}(l_0=\vert m\vert+6)&\cdots\\
i_{1\phm{-}}(l_0=\vert m\vert+2)&i_{1\phm{-}}(l_0=\vert m\vert+4)&
i_{1\phm{-}}(l_0=\vert m\vert+6)&\cdots\\
i_{-1}(l_0=\vert m\vert+2)      &i_{-1}(l_0=\vert m\vert+4)    &
i_{-1}(l_0=\vert m\vert+6)&\cdots\\
                                &i_{-2}(l_0=\vert m\vert+4)      &
i_{-2}(l_0=\vert m\vert+6)&\cdots\\
                              & &i_{-3}(l_0=\vert m\vert+6)&\cdots\\
 & & &\cdots\\
\end{array} 
\right]& \label{ina}\\
& & \nonumber\\
&\Updownarrow& \nonumber\\
& & \nonumber\\
&\left[
\begin{array}{llll}
 & & &\cdots\\
 & &g_{1\phm{-}}(l_{\phm{0}}=\vert m\vert+4) &\cdots\\ 
 &g_{1\phm{-}}(l_{\phm{0}}=\vert m\vert+2)&
  g_{2\phm{-}}(l_{\phm{0}}=\vert m\vert+2)&\cdots\\
g_{1\phm{-}}(l_{\phm{0}}=\vert m\vert)\phm{+2}& 
g_{2\phm{-}}(l_{\phm{0}}=\vert m\vert)\phm{+4}&
g_{3\phm{-}}(l_{\phm{0}}=\vert m\vert)\phm{+6}&\cdots\\
g_{-1}(l_{\phm{0}}=\vert m\vert)\phm{+2}      & 
g_{-2}(l_{\phm{0}}=\vert m\vert)\phm{+4}      &
g_{-3}(l_{\phm{0}}=\vert m\vert)\phm{+6}      &\cdots\\
 &g_{-1}(l_{\phm{0}}=\vert m\vert+2)      &
  g_{-2}(l_{\phm{0}}=\vert m\vert+2)      &\cdots\\
 & &g_{-1}(l_{\phm{0}}=\vert m\vert+4) &\cdots\\
 & & &\cdots\\
\end{array}
\right]& \ , \label{gra}
\end{eqnarray}
where the two corresponding modes occupy the same position 
in the two matrices connected with an arrow.

From the diagrams,
we may find the connection rule between the $g_{\pm k}$-modes with $l$ and $m$ and 
the inertial modes $i_{\pm j}$ with $l_0$ and $m$: 
\begin{equation}
g_{\pm k}(l=l^j)\longleftrightarrow  i_{\pm j}(l_0=l^j+2k), \label{g-i}
\end{equation}
where $l^j=|m|+2j-2$ for even modes and $l^j=|m|+2j-1$ for odd modes,
and $j$ and $k$ are positive integers.
The symbol $g_{\pm k}$ denotes the $g$-mode
with $k$ radial nodes of the expansion coefficient $S_l$, 
having positive and negative frequencies $\omega$
observed in the corotating frame of the star when $\bar\Omega\not=0$.

The connection rule between the $r$-modes
associated with $l'=|m|$ and the inertial modes $i_0$ with $l_0$ and $m$ may be given by
\begin{equation}
r_{j-1} ( l' = \vert m \vert ) \longleftrightarrow i_0 
(l_0 = |m| + 2 j -1) \, ,
\label{r-i-odd}
\end{equation}
where $j$ is a positive integer.
Note that the 
$i_0$-mode with $l_0 = \vert m \vert +1$ is exactly the same as 
the $r_0$-modes with $l'=\vert m \vert$. 
To establish the connection rules given above, we have neglected the effects of
avoided crossings between the modes associated with different values of $l$ and/or $k$, since
the mode property is carried over at the crossings.

%%%%%%%%%%%%%%%%%%%%%%%%%%%%%%%%%%%%%%%%%%%%%%%%%%%%%%%%%%%%%%
%% V. DISSIPATION TIMESCALES .....                          %%
%%%%%%%%%%%%%%%%%%%%%%%%%%%%%%%%%%%%%%%%%%%%%%%%%%%%%%%%%%%%%%

\section{Dissipation Timescales for $r$-modes of Non-isentropic Stars}

For the rotational modes of non-isentropic models, we have found that
all the $r$-modes except the $r_0$-mode with $l'=|m|=1$ 
are unstable to the gravitational radiation reaction 
in the sense that the frequency $\sigma$ satisfies the condition 
$\sigma(\sigma+m\Omega)<0$.
Let us determine the stability of the $r$-modes of the non-isentropic models
against the gravitational radiation driven instability, taking account of
the dissipative processes due to the shear and bulk viscosity, where
we will follow the method of calculation almost the same as that employed
in the previous stability analyses of the $r_0$-mode with $l'=|m|$ and the
inertial modes for isentropic models
(Lindblom et al 1998, Owen et al 1998, Andersson et al 1998, 
Kokkotas \& Stergioulas 1998, Lindblom et al 1999, Lockitch \& Friedman 1999,
Yoshida \& Lee 2000).

The damping timescale of the $r_0$-mode associated with $l'=|m|$
for sufficiently small $\bar{\Omega}$ may be estimated by (Lindblom et al 1998)
\begin{eqnarray}
\frac{1}{\tau} &=& \frac{1}{\tilde \tau_S} \left( \frac{10^9 K}{T} \right)^2
+ \frac{1}{\tilde \tau_B} \left( \frac{T}{10^9 K} \right)^6
\left( \frac{\Omega^2}{\pi G \bar{\rho}} \right) \nonumber \\
&+& \sum_{l=2}^{\infty} \frac{1}{\tilde \tau_{J,l}} 
\left( \frac{\Omega^2}{\pi G \bar{\rho}} \right)^{l+1}
+ \sum_{l=2}^{\infty} \frac{1}{\tilde \tau_{D,l}} 
\left( \frac{\Omega^2}{\pi G \bar{\rho}} \right)^{l+2} \, ,
 \label{tau2}
\end{eqnarray}
where $\bar{\rho}$ is the average density of the star.
The dissipative timescales $\tilde\tau$ are calculated for each oscillation mode
in the limit of $\bar\Omega\rightarrow 0$ so that
they are independent of the rotation frequency $\Omega$.
Here, the first, second, third and fourth terms in the right-hand side of 
equation (\ref{tau2}) are contributions from the shear viscosity, 
the bulk viscosity, the current multipole radiation and the mass 
multipole radiation, respectively. 
The expression (\ref{tau2}) of the damping timescales has been derived on the assumptions that
the frequency of the modes is well represented by a linear function of $\Omega$ and the
toroidal component of the displacement vector is dominating.
Although the assumptions are reasonable for the $r_0$-modes with $l'=|m|$ 
(see Yoshida and Lee 2000),
they are not necessarily correct for the $r_k$-modes with $l'=|m|$ and $k>0$, 
for which we use
\begin{eqnarray}
\frac{1}{\tau} &=& \frac{1}{\tilde \tau_S} \left( \frac{10^9 K}{T} \right)^2
+ \frac{1}{\tilde \tau_B} \left( \frac{T}{10^9 K} \right)^6
\left( \frac{\pi G \bar{\rho}}{\Omega^2} \right) \nonumber \\
&+& \sum_{l=2}^{\infty} \frac{1}{\tilde \tau_{J,l}} 
\left( \frac{\Omega^2}{\pi G \bar{\rho}} \right)^{l+1}
+ \sum_{l=2}^{\infty} \frac{1}{\tilde \tau_{D,l}} 
\left( \frac{\Omega^2}{\pi G \bar{\rho}} \right)^{l+2} \, ,
 \label{tau2b}
\end{eqnarray}
because these $r$-modes become inertial modes at large $\bar\Omega$.
The difference between these two expressions is found
in the second terms in the right-hand side of the two equations. 
When $\bar\Omega$ is small, equation (\ref{tau2b}) gives the correct 
expression of $\tau^{-1}$ for inertial modes of isentropic stars
(See, Lockitch \& Friedman 1999, Yoshida \& Lee 2000).
To determine the damping timescales $\tilde\tau$ in (\ref{tau2b}) in the limit of 
$\bar\Omega\rightarrow 0$,
we assume that the dissipative timescale $\tilde\tau^{-1}$ 
depends on $\bar\Omega$ as $a\bar\Omega^2+b$ for each dissipative process, where
the two unknown constants $a$ and $b$ are
determined by calculating $\tilde\tau^{-1}$ and $d\tilde\tau^{-1}/d\bar\Omega$
at $\bar\Omega=0.1$.
The $\tilde\tau^{-1}$ in the limit of $\bar\Omega\rightarrow0$ is given by
$\tilde\tau^{-1}=b$.
We note that equation (\ref{tau2b}) is not necessarily a correct approximation for 
non-isentropic stars. 
We use the expression (\ref{tau2b}) simply 
because we want to give rough 
estimates of the dissipative timescales, in order to compare the timescales 
with those of the isentropic case.

In Table 3, we tabulate the dissipation timescales $\tilde\tau$ in the unit of second for 
the various dissipative processes for the $r_k$-modes with $l'=\vert m \vert$, 
where the radius and the mass of the $n=1$ polytropic neutron star model
at $\Omega=0$ are chosen to be $R=12.57\mbox{km}$ and $M=1.4M_{\sun}$, 
respectively. 
Note that in Table 3 we have not included the $r_k$-modes with $l'>|m|$,
since the frequency $\omega$ becomes vanishingly small at large $\bar\Omega$ 
(see Figures 1 to 3) 
and hence the magnitude of the gravitational radiation driven
instability, which is proportional to $\omega$, becomes quite small.
For the $r_0$-modes with $l'=|m|$,
in order to see the effects of the buoyancy force on 
the stability, the results for the four different values of $\gamma$ are tabulated. 
As shown by Table 3, the gravitational radiation driven instability 
for the $r_0$-modes with $l'=\vert m \vert$ 
is not affected by introducing the deviation from isentropic structure and remains
strong even for the non-isentropic models.
As for the $r_k$-modes with $l'=|m|$ and $k> 0$, it is found that 
the dissipative timescales are comparable to those of the corresponding inertial 
modes (compare Table 4 of Yoshida \& Lee 2000). Thus, we may conclude that the $r$-mode 
instability driven by the gravitational radiation reaction is dominated by 
the $r_0$-mode with $l'=\vert m \vert = 2$ both in isentropic and in non-isentropic stars.

%%%%%%%%%%%%%%%%%%%%%%%%%%%%%%%%%%%%%%%%%%%%%%%%%%%%%%%%%%%%%%
%%       CONCLUSIONS                                        %%
%%%%%%%%%%%%%%%%%%%%%%%%%%%%%%%%%%%%%%%%%%%%%%%%%%%%%%%%%%%%%%

\section{Conclusions}

In this paper, we have numerically investigated the properties of rotational 
modes for slowly rotating, 
non-isentropic, Newtonian stars, where the effects of 
the centrifugal force are taken into consideration to calculate the modes
with accuracy up to the order of $\bar\Omega^3$.
Estimating the dissipation timescales due to the 
gravitational radiation and the shear and bulk viscosity
for simple polytropic neutron star models with the index $n=1$,
we have shown that the gravitational radiation driven instability
of the $r_0$-mode with $l'=|m|$ is hardly 
affected by introducing the deviation from the
isentropic structure, and that the instability of the $r_0$-modes with $l'=|m|=2$
remains strongest even for the non-isentropic models.
Note that the $r_0$-modes with $l'=|m|$ are the only $r$-modes possible in
isentropic stars and their modal property does not very much depend on $\Omega$ and 
the deviation from isentropic structure.
We thus conclude that 
the result of the stability analysis of the $r_0$-modes with $l'=|m|$
for isentropic stars remains valid even for non-isentropic stars.

We have numerically found that for the radiative models
the inertial modes at large $\bar\Omega$ 
become identical with the $r_k$-modes with $l'=|m|$ and $k>0$ and $g$-modes
in the limit of $\bar\Omega\rightarrow 0$ because of the effect of the buoyant force, 
where the integer $k$ denotes the number of node of the 
eigenfunction with dominant amplitude.
This behavior of the $g$-modes or the inertial modes as functions of $\Omega$
has been suggested by, for example,
Saio (1999) and  Lockitch \& Friedman (1999).
We find the connection rules between the $r_k$-modes with $l'=|m|$ and $k>0$ and
$g$-modes at small $\bar\Omega$ and
the inertial modes at large $\bar\Omega$.
We also suggest that the $r_k$-modes with $l'>|m|$ and $k\ge 0$, 
whose frequency in the corotating frame
becomes vanishingly small as $\bar\Omega$ increases, 
have at large $\bar\Omega$ no corresponding inertial
modes of the isentropic models.

Good progress has been made in the understanding of
the properties of the $r$-mode instability 
due to the gravitational radiation since its discovery.
Recently, for example, Yoshida et al (1999) have shown that the results 
of the $r$-mode instability obtained by using the slow rotation approximation,
as employed in this paper,
are consistent with the results obtained 
by calculating directly the $r$-mode oscillations 
of rotating stars without the approximation. 
Kojima (1998) and Kojima \& Hosonuma (1999) have studied the time evolutional 
behavior of $r$-modes by using the Laplace transformation technique
for slowly rotating relativistic models. 
Lockitch (1999) also obtained $r$-modes and   
rotation modes for slowly rotating, incompressible, relativistic stars. 
However, we believe it necessary to show
that the gravitational radiation driven instability
of $r$-modes remains sufficiently robust in more realistic and unrestricted situations,
taking account of the effects of differential rotation, magnetic field 
and general relativity, for example.
To investigate the $r$-mode instability in more general situations is one of
our future studies.

%%%%%%%%%%%%%%%%%%%%%%%%%%%%%%%%%%%%%%%%%%%%%%%%%%%%%%%%%%%%%%
%%         END OF MAIN BODY OF PAPER                        %%
%%%%%%%%%%%%%%%%%%%%%%%%%%%%%%%%%%%%%%%%%%%%%%%%%%%%%%%%%%%%%%

\acknowledgements

We would like to thank Prof. H. Saio for useful comments and discussions.  
S.Y. was supported by Research Fellowship of the Japan Society for 
the Promotion of Science for Young Scientists.

%%%%%%%%%%%%%%%%%%%%%%%%%%%%%%%%%%%%%%%%%%%%%%%%%%%%%%%%%%%%%%
%%  BIBLIOGRAPHY                                            %%
%%%%%%%%%%%%%%%%%%%%%%%%%%%%%%%%%%%%%%%%%%%%%%%%%%%%%%%%%%%%%%

%

%%%%%%%%%%%%%%%%%%%%%%%%%%%%%%%%%%%%%%%%%%%%%%%%%%%%%%%%%%%%%%
%% FIGURES                                                  %%
%%%%%%%%%%%%%%%%%%%%%%%%%%%%%%%%%%%%%%%%%%%%%%%%%%%%%%%%%%%%%%

\begin{figure}
\epsscale{.6}
\plotone{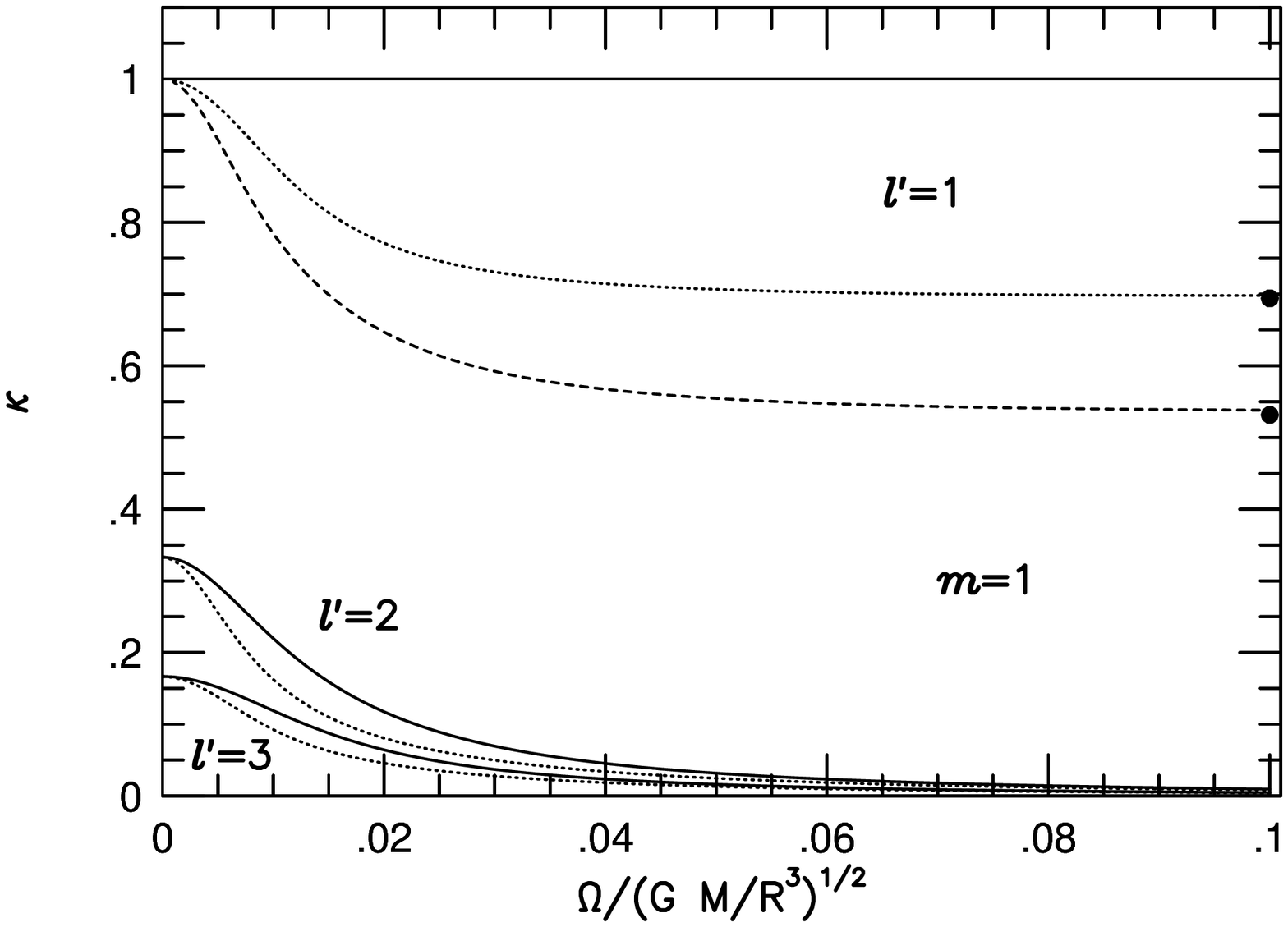}
\caption{Dimensionless frequencies $\kappa=\omega/\Omega$ for the
low-radial-order $r$-modes with $m=1$ are plotted as functions of the dimensionless 
angular rotation frequency $\bar{\Omega}=\Omega/(GM/R^3)^{1/2}$, 
for the $n=1$ polytropic model with $\gamma=-10^{-4}$. 
The solid, the dotted and the dashed curves are
for the $r_0$, $r_1$, and $r_2$-modes, respectively. 
The attached label $l'$ 
denotes the angular quantum number $l'$ associated with the spherical harmonic function
representing the toroidal component of the displacement vector.
Note that the eigenvalue of the 
$r$-modes is given by $\kappa_0=2 m/(l'(l'+1))$ in the limit of $\bar\Omega\rightarrow 0$. 
The filled circles at $\bar{\Omega}=0.1$ denote the eigenvalues $\kappa$ of 
the inertial modes calculated for the isentropic model 
with the index $n=1$ and $\gamma=0$ at $\bar{\Omega}=0.1$. }
\end{figure}

\begin{figure}
\epsscale{.6}
\plotone{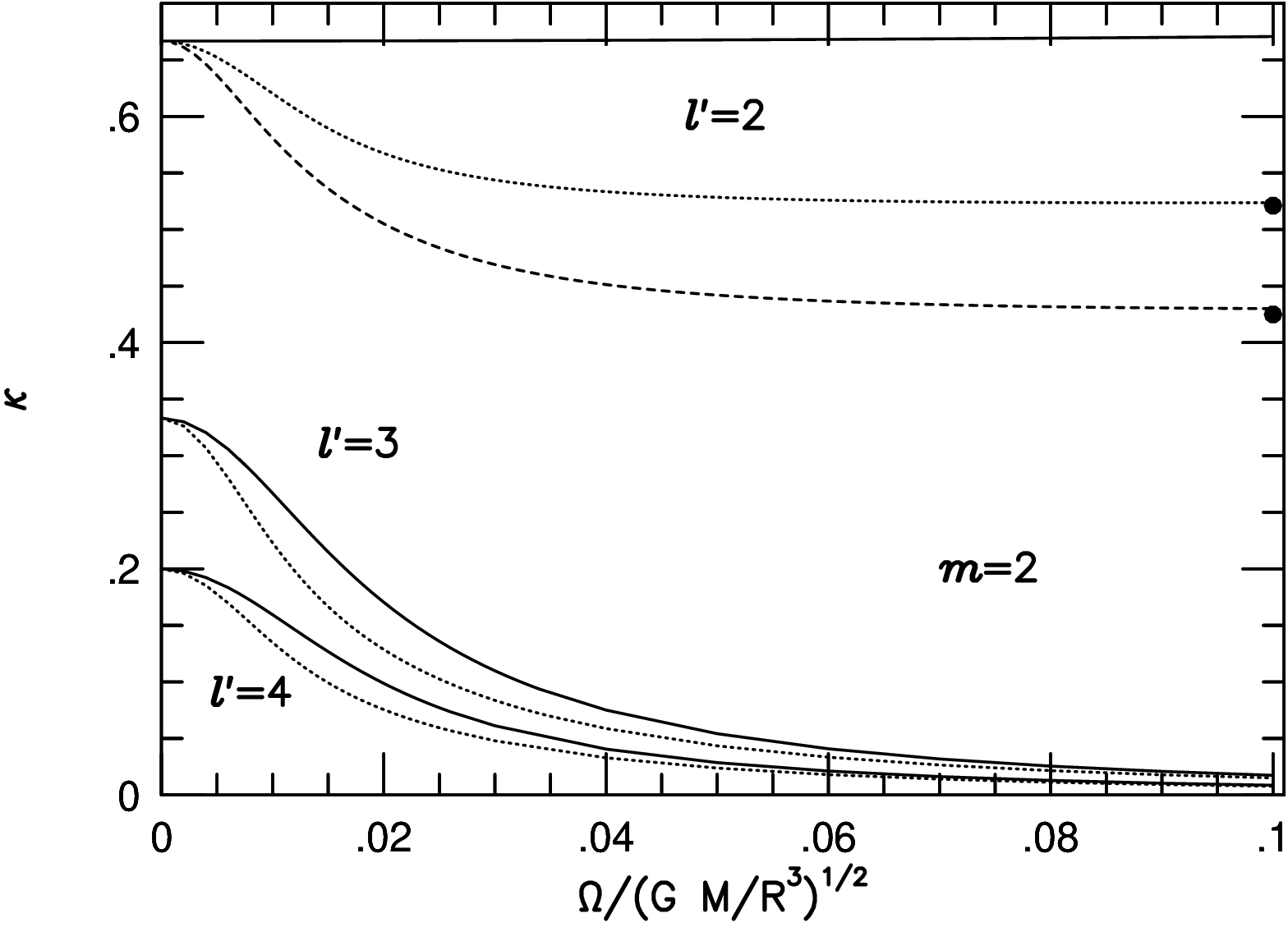}
\caption{Same as Figure 1, but for $m=2$.}
\end{figure}

\begin{figure}
\epsscale{.6}
\plotone{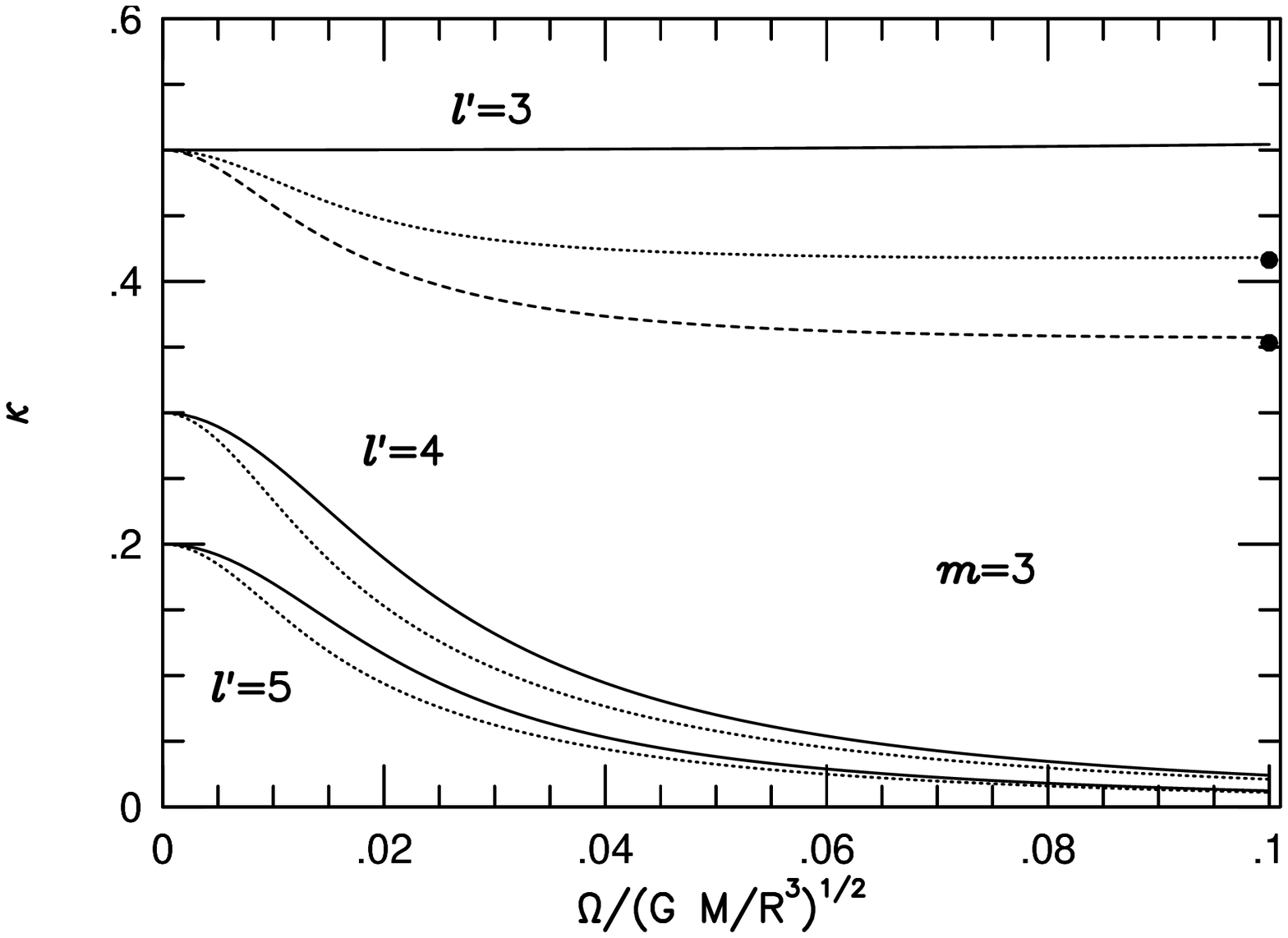}
\caption{Same as Figure 2, but for $m=3$.}
\end{figure}

\begin{figure}
\epsscale{.6}
\plotone{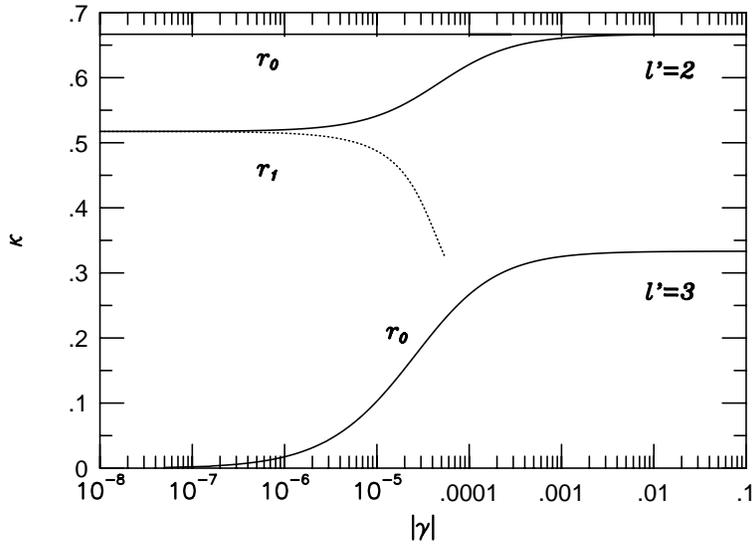}
\caption{Dimensionless frequencies $\kappa$ for the low-radial-order 
$r$-modes with $m=2$ are plotted as functions of $|\gamma|$ for the $n=1$ polytropic model 
at $\bar{\Omega}=10^{-2}$. 
The attached label $l'$ denotes the index of the spherical harmonic function for 
the toroidal component of the  displacement vector.
The number of nodes of $iT_{l'}$ in the radial direction 
is indicated by the subscript in the notation $r_k$ attached to the mode sequences. 
The solid line and the dotted line are given for negative 
and positive values of $\gamma$, respectively.} 
\end{figure}

\begin{figure}
\epsscale{.4}
\plottwo{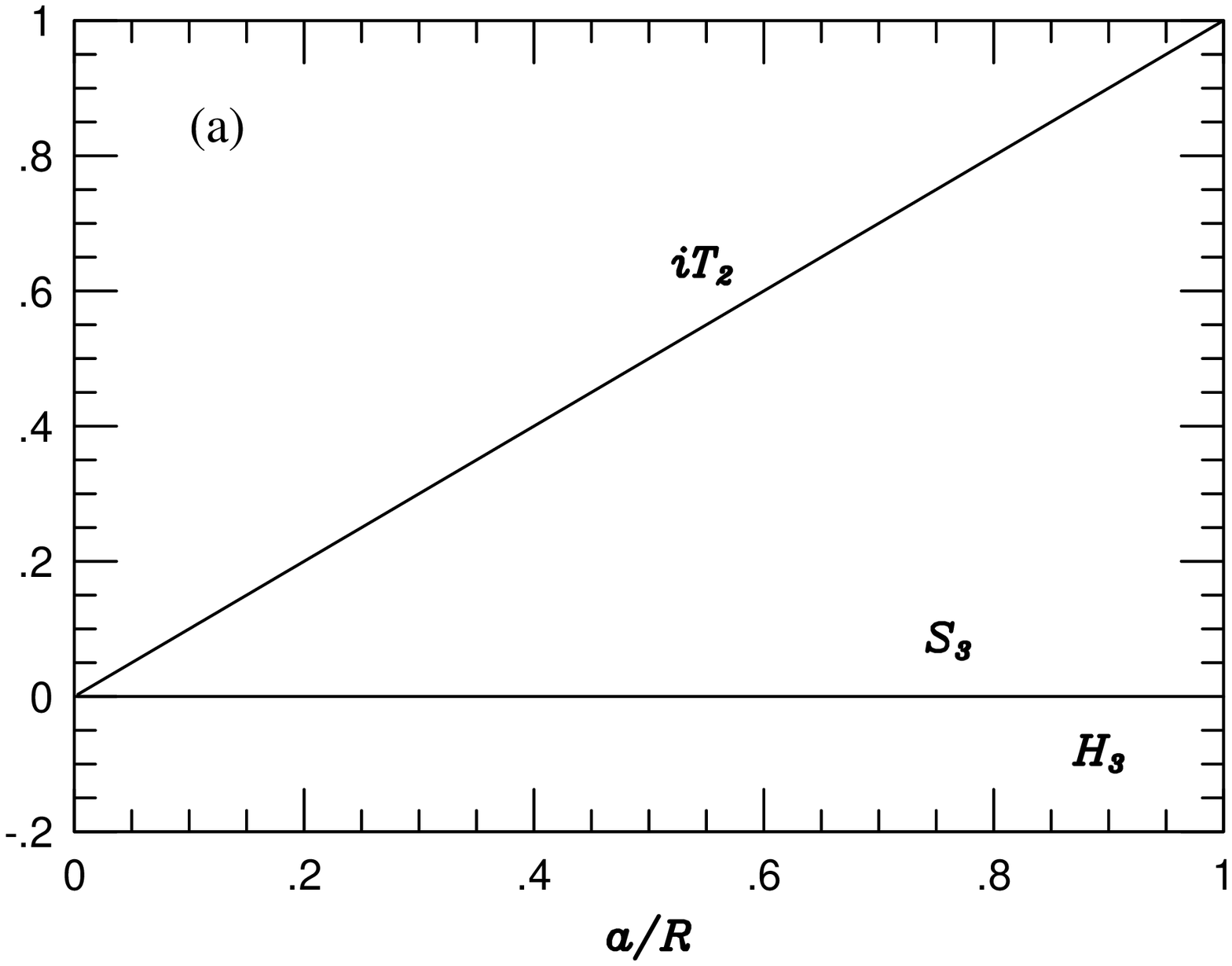}{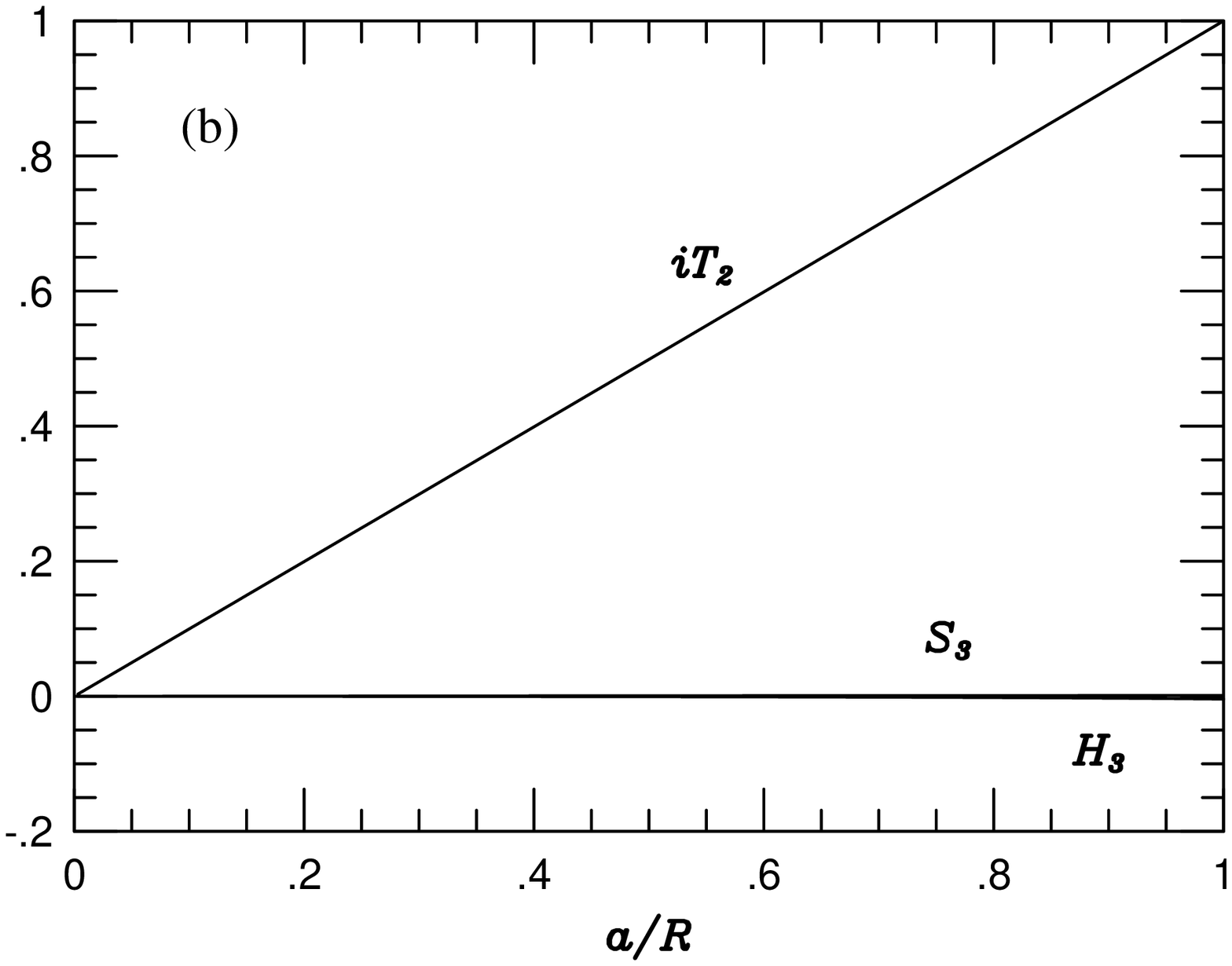}
\caption{The first expansion coefficients $S_l$, $H_l$, $iT_{l'}$ for 
the $r_0$ mode with $l'=m=2$ are plotted against $a/R$ for 
the polytropic model with the index $n=1$ and $\gamma=-10^{-4}$. 
Panels $(a)$ and $(b)$ are for the cases of
$\bar{\Omega}=10^{-3}$ and $\bar{\Omega}=10^{-1}$, respectively. 
The eigenfunctions are normalized so that $T_{l'}(a=R)=1$. 
The attached labels $S_l$, $H_l$, and $iT_{l'}$ denote the expansion coefficients 
of the displacement vector.}
\end{figure}

\begin{figure}
\epsscale{.4}
\plottwo{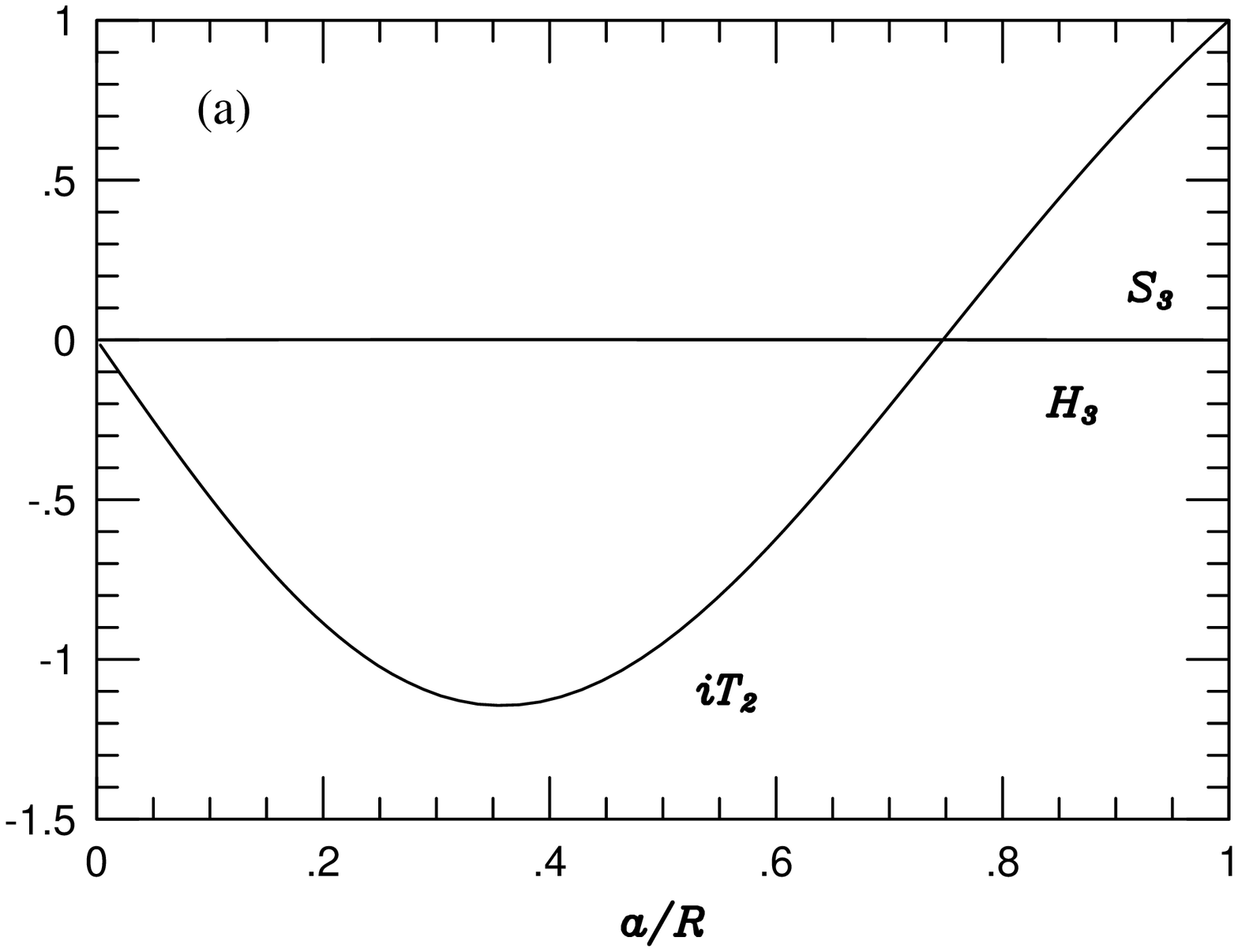}{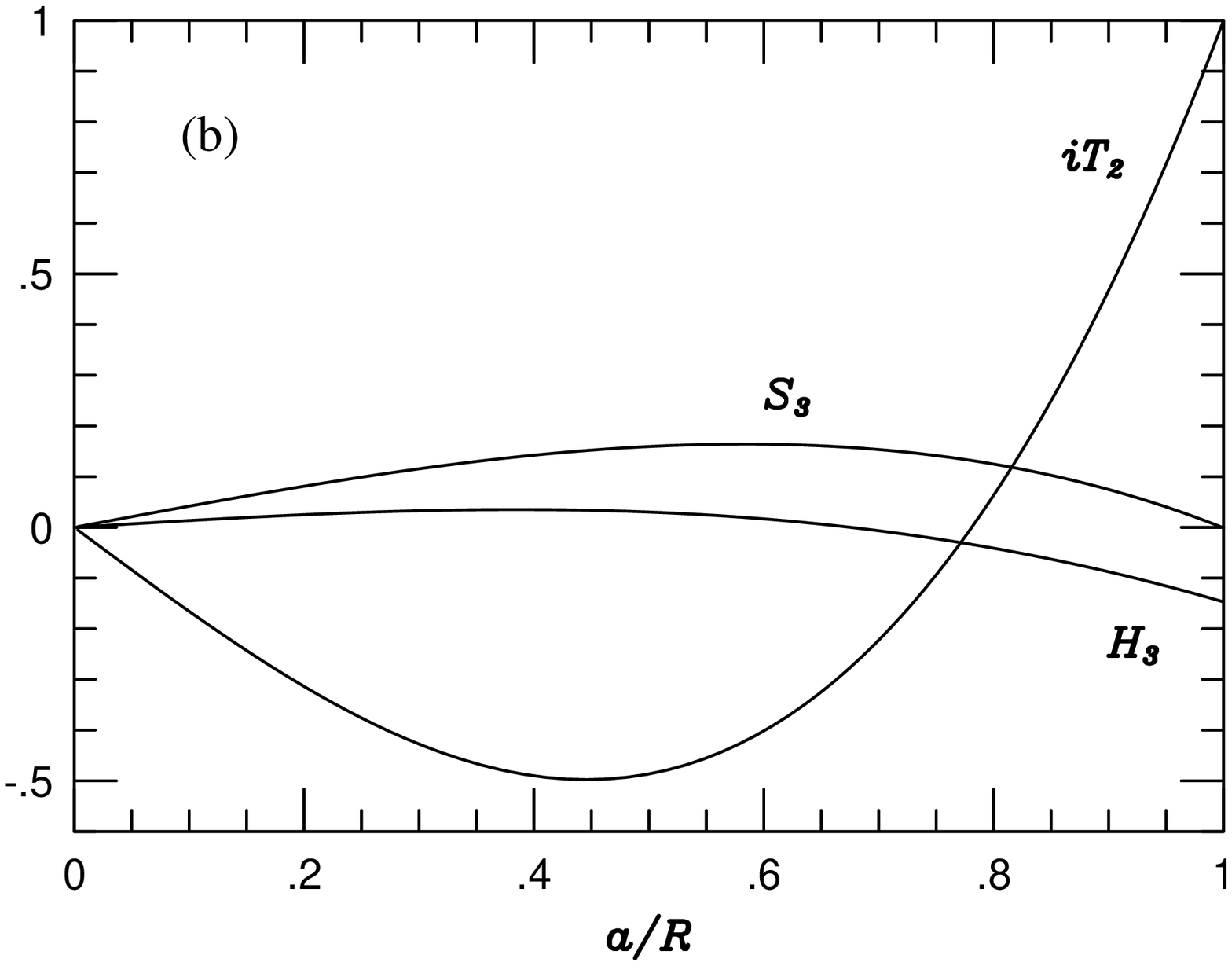}
\caption{Same as Figure 5, but for the $r_1$-mode with $l'=2$.}
\end{figure}

\begin{figure}
\epsscale{.4}
\plottwo{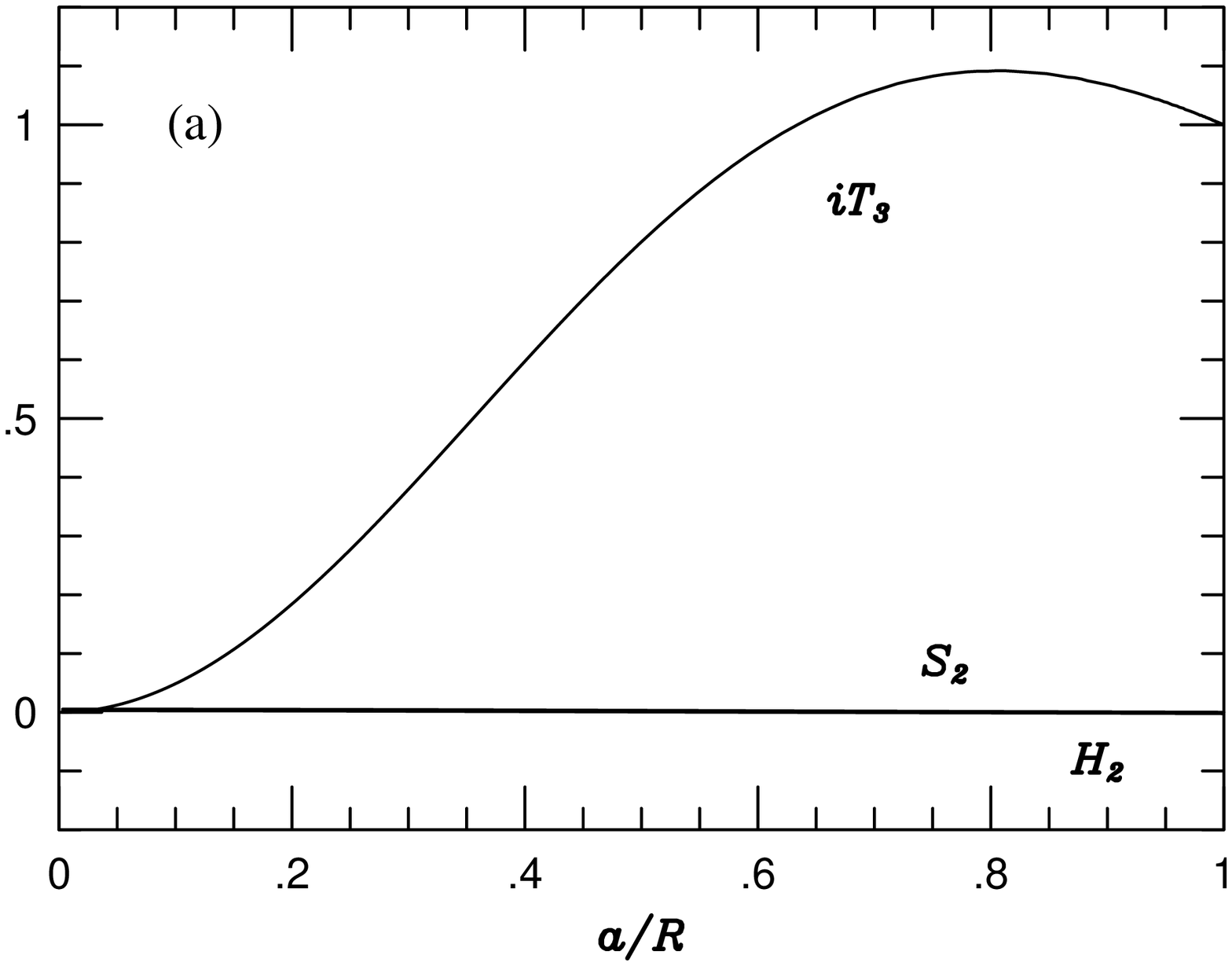}{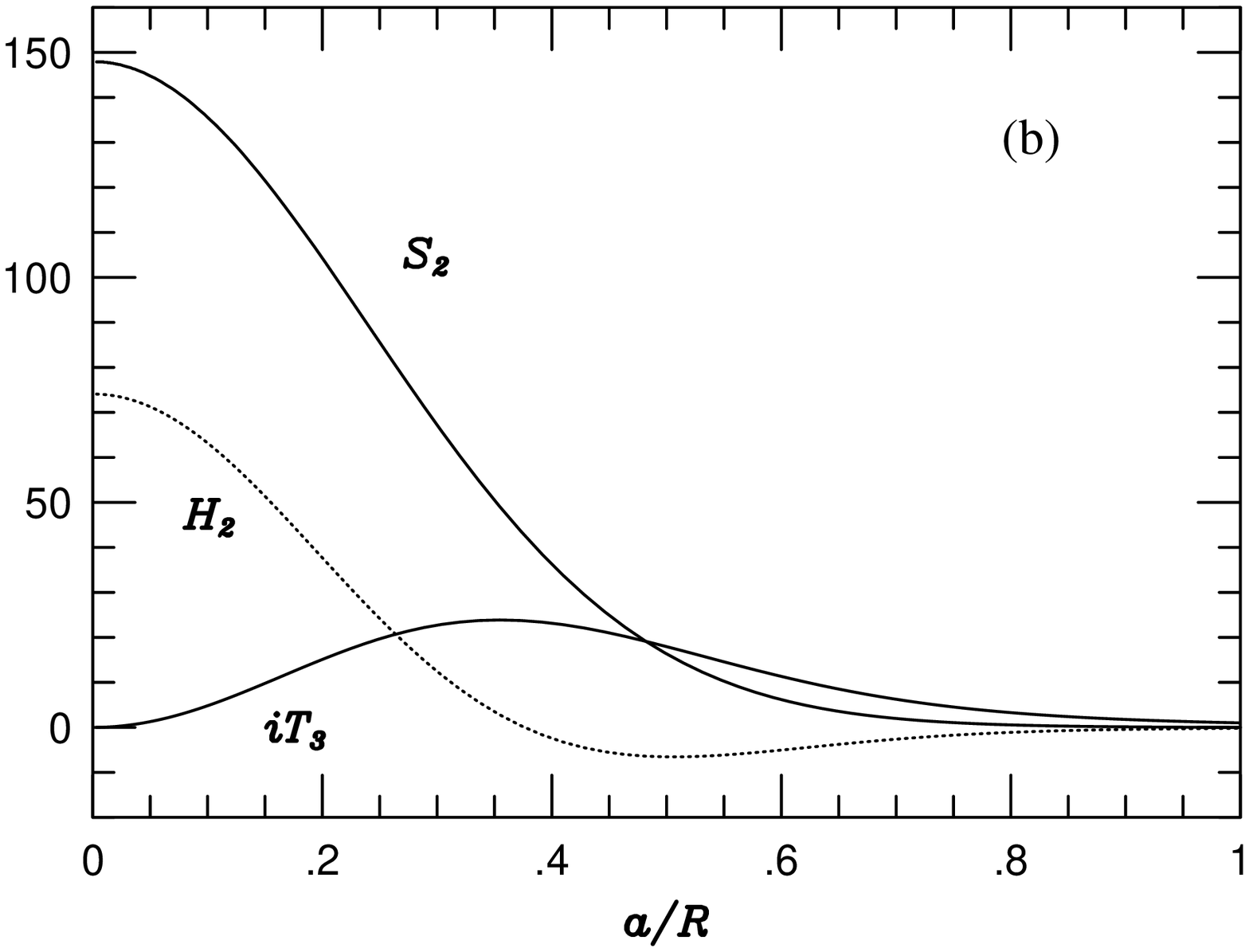}
\caption{Same as Figure 5, but for the $r_0$-mode with $l'=3$.}
\end{figure}

\begin{figure}
\epsscale{.4}
\plottwo{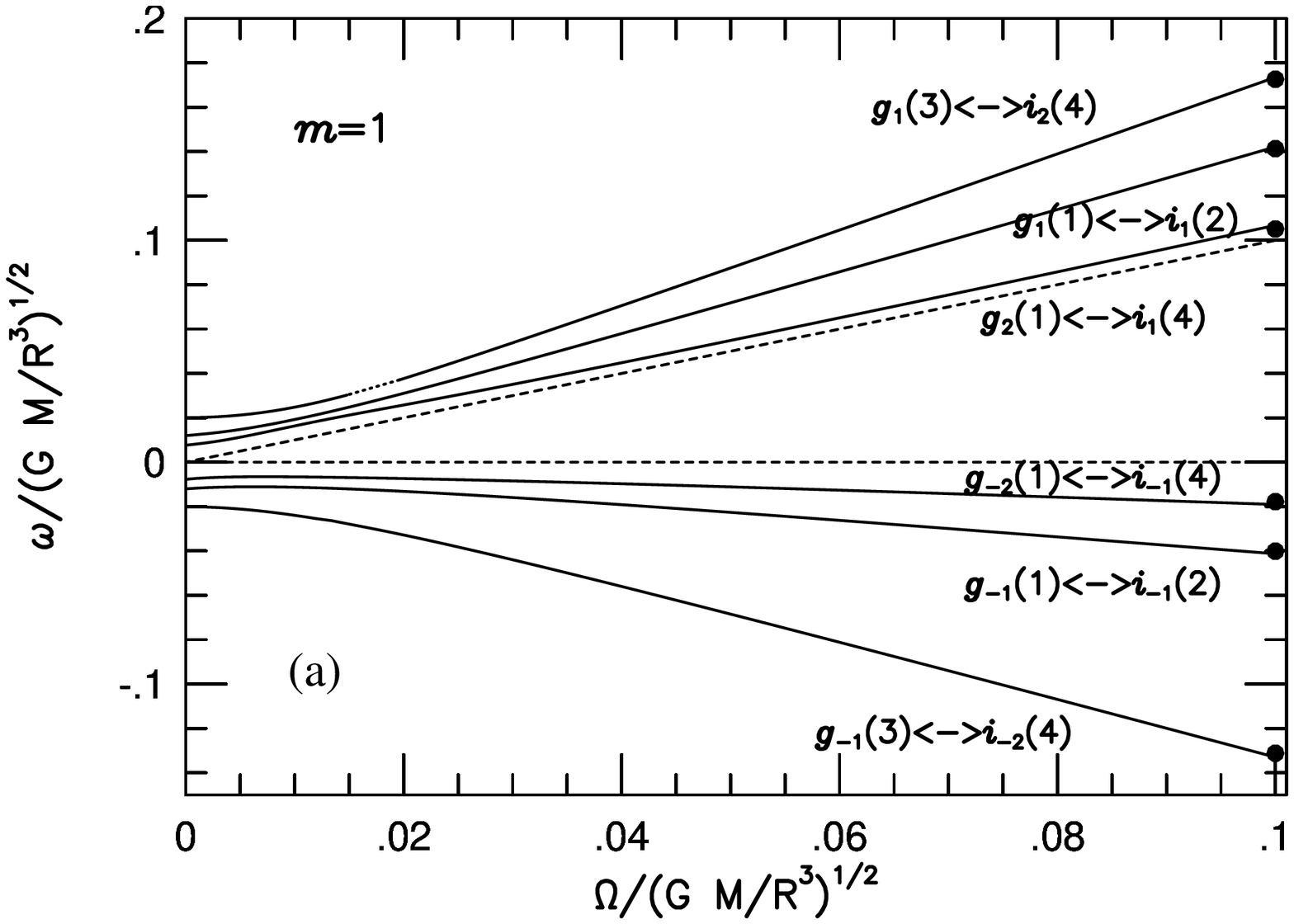}{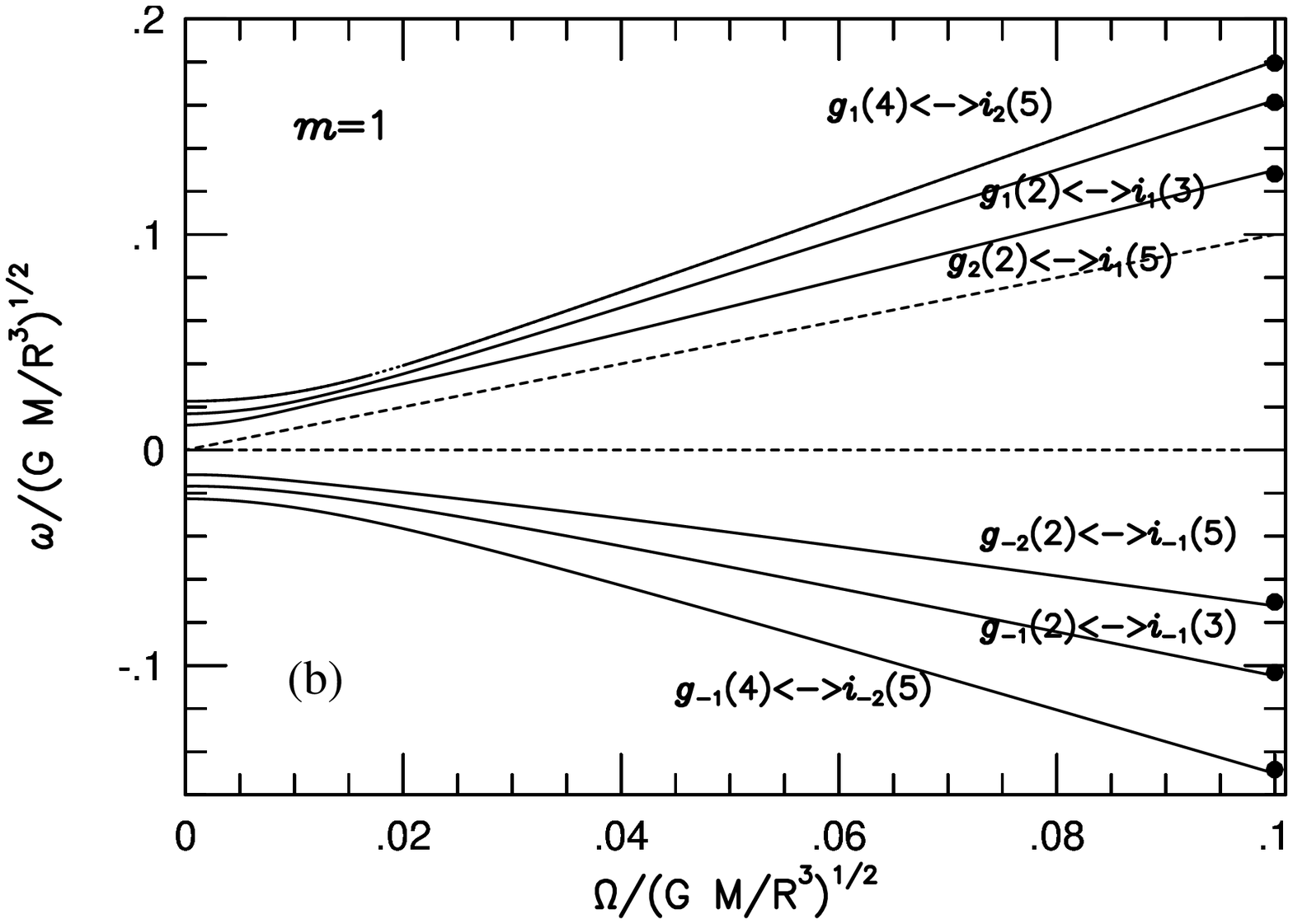}
\caption{Dimensionless frequencies 
$\bar{\omega}=\omega/(GM/R^3)^{1/2}$ for the low-radial-order 
$g$- and inertial modes with $m=1$ are plotted as functions of 
$\bar{\Omega}=\Omega/(GM/R^3)^{1/2}$, 
for the $n=1$ polytropic model with $\gamma=-10^{-4}$. 
The even and the odd parity modes are shown in panels $(a)$ and $(b)$, 
respectively.  
The attached labels 
$g_{\pm k}(l)\longleftrightarrow i_{\pm j}(l_0-\vert m\vert)$ 
indicates the connection rule between the $g$-modes and inertial modes.
The filled circles at $\bar{\Omega}=0.1$ denote the eigenfrequencies $\bar\omega$
of the inertial modes for the $n=1$ isentropic model at $\bar{\Omega}=0.1$.
The part indicated by the dotted line in the mode sequence is the domain
of $\bar\Omega$, in which the modes belonging to the sequence cannot be 
calculated. The two dashed lines are given by $\bar{\omega}=0$ 
and $\bar{\omega}= m \bar{\Omega}$.} 
\end{figure}

\begin{figure}
\epsscale{.4}
\plottwo{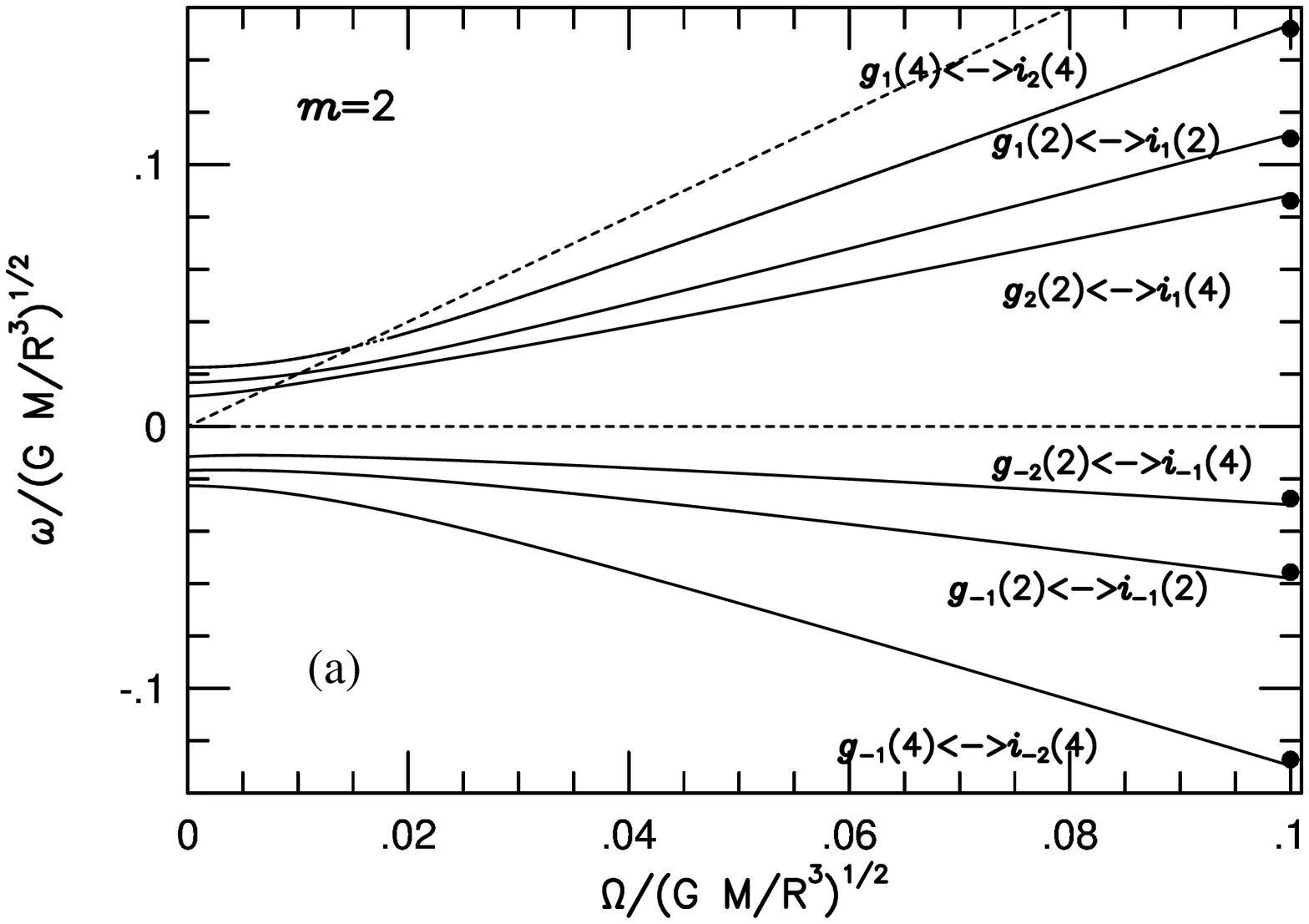}{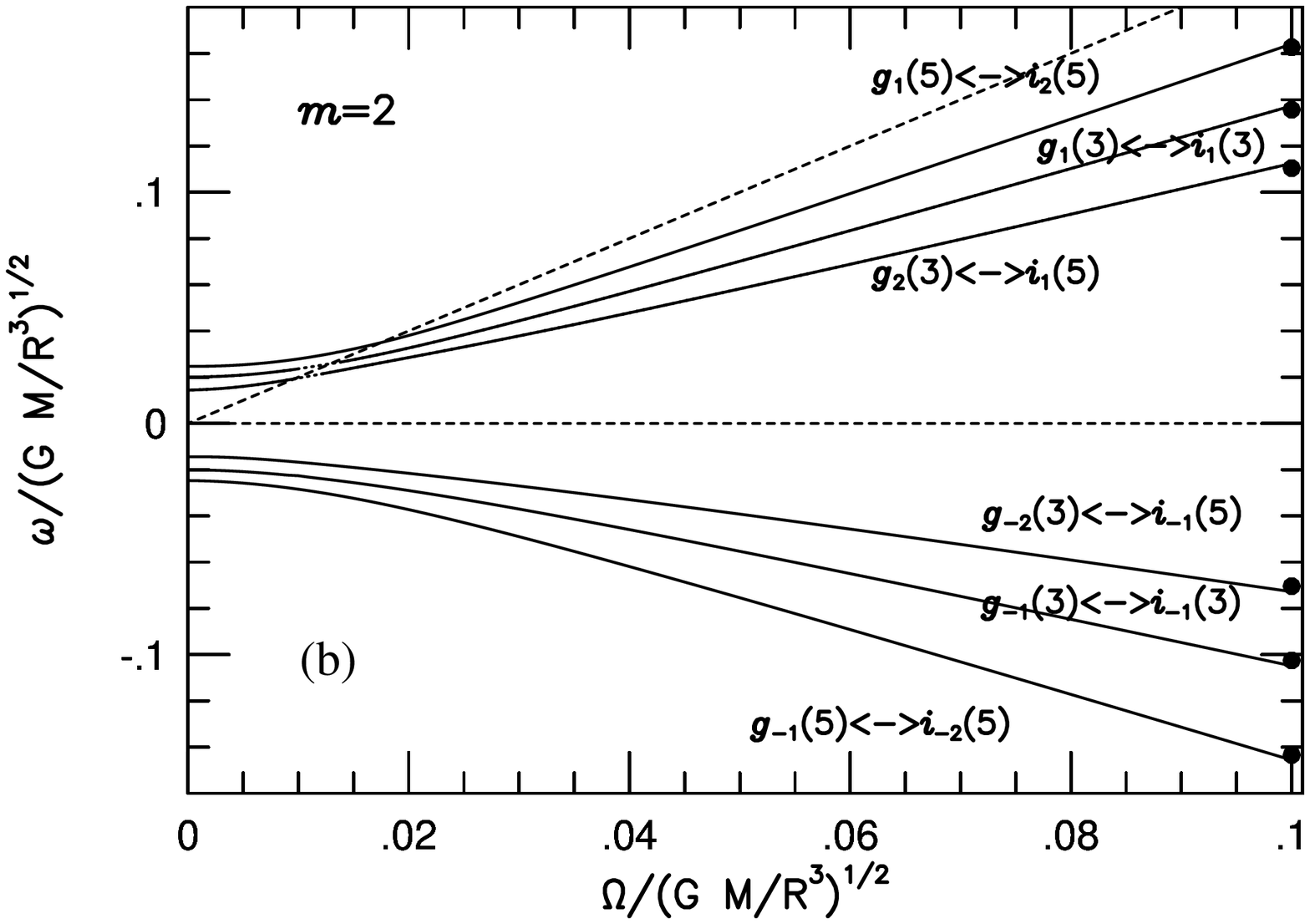}
\caption{Same as Figure 8, but for $m=2$.}
\end{figure}

\begin{figure}
\epsscale{.4}
\plottwo{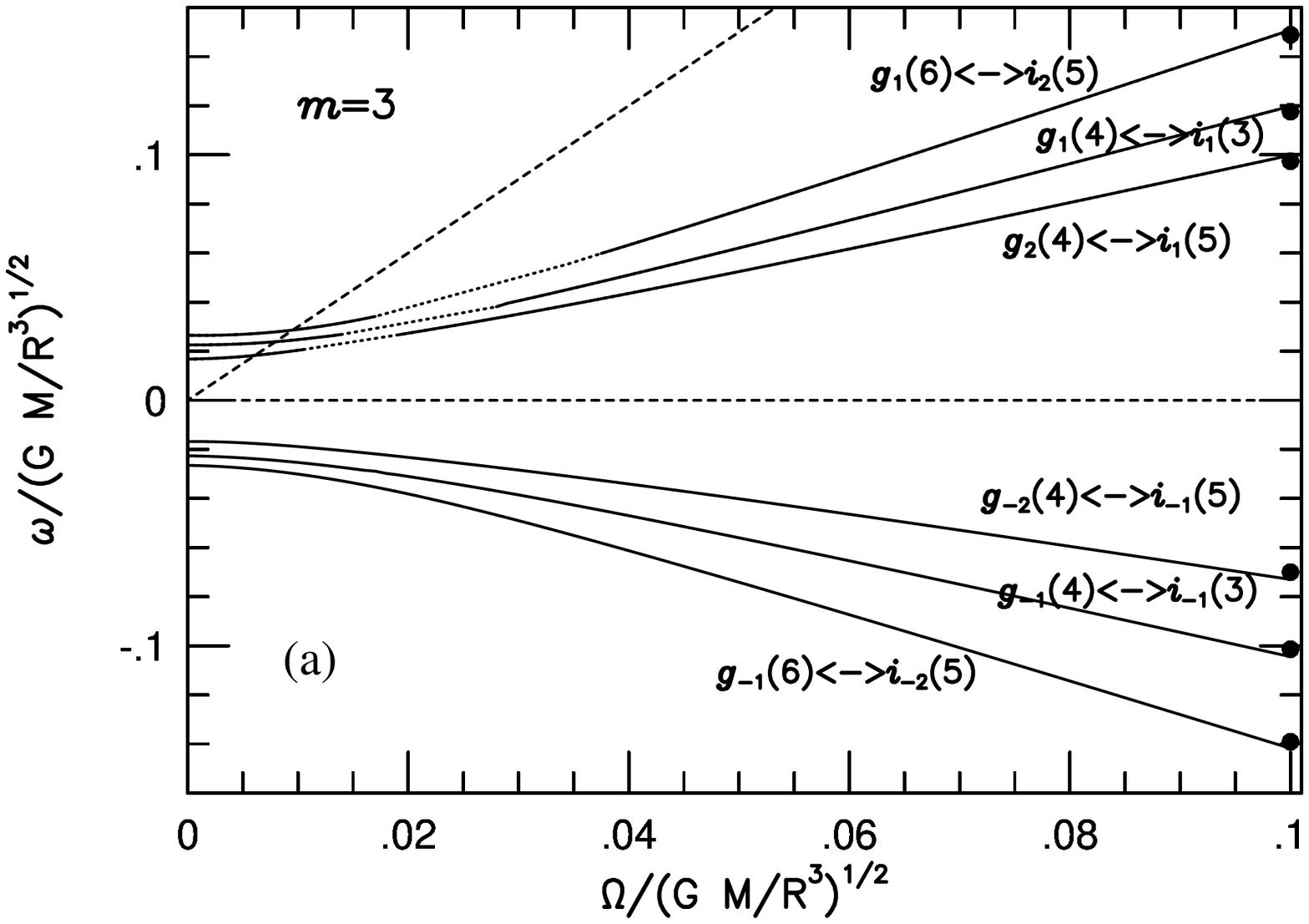}{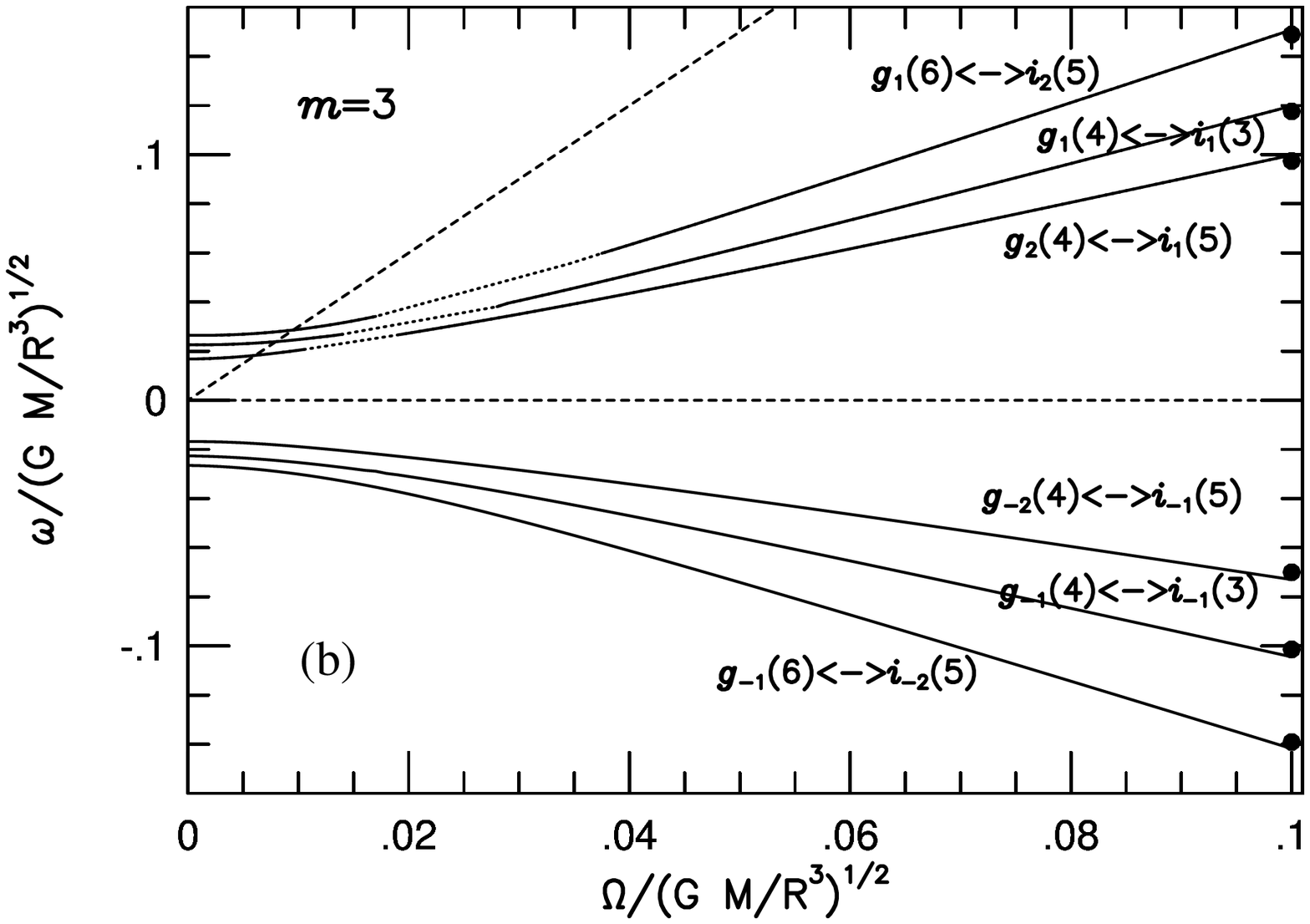}
\caption{Same as Figure 9, but for $m=3$.}
\end{figure}

%%%%%%%%%%%%%%%%%%%%%%%%%%%%%%%%%%%%%%%%%%%%%%%%%%%%%%%%%%%%%%
%% TABLES                                                   %%
%%%%%%%%%%%%%%%%%%%%%%%%%%%%%%%%%%%%%%%%%%%%%%%%%%%%%%%%%%%%%%

\clearpage

\begin{deluxetable}{ccccc}
\footnotesize
\tablecaption{$(\omega/\Omega-\kappa_0)/{\bar{\Omega}}^2\ 
$\tablenotemark{a} \ 
for $r$-modes of the $n=3$ polytrope with $\Gamma=5/3$ \label{code-check}}
\tablewidth{250pt}
\tablehead{
\colhead{$m$} & \colhead{$l'$} & \colhead{Mode} & 
\colhead{Present}  & \colhead{Saio 1982}  
}
\startdata
 1 & 1 & $r_0$ & \phm{-}0.000 & \phm{-}0.000  \nl
   &   & $r_1$ &       -0.040 &       -0.040  \nl
   &   & $r_2$ &       -0.115 &       -0.114  \nl
   & 2 & $r_0$ &       -0.100 &       -0.099  \nl
   &   & $r_1$ &       -0.191 &       -0.190  \nl
   &   & $r_2$ &       -0.315 &       -0.312  \nl
   & 3 & $r_0$ &       -0.037 &       -0.037  \nl
   &   & $r_1$ &       -0.063 &       -0.061  \nl
   &   & $r_2$ &       -0.097 &       -0.096  \nl
 2 & 2 & $r_0$ & \phm{-}0.121 & \phm{-}0.120  \nl
   &   & $r_1$ & \phm{-}0.024 & \phm{-}0.024  \nl
   &   & $r_2$ &       -0.001 &       -0.001  \nl
   & 3 & $r_0$ &       -0.037 &       -0.037  \nl
   &   & $r_1$ &       -0.071 &       -0.070  \nl
   &   & $r_2$ &       -0.115 &       -0.114  \nl
 3 & 3 & $r_0$ & \phm{-}0.174 & \phm{-}0.173  \nl
   &   & $r_1$ & \phm{-}0.064 & \phm{-}0.063  \nl
   &   & $r_2$ & \phm{-}0.028 & \phm{-}0.027  \nl
\enddata
\tablenotetext{a}{$\kappa_0$ is defined as $\kappa_0=2 m/(l' (l'+1))$. 
The eigenfrequencies are calculated at $\bar{\Omega}=0.1$. }
\end{deluxetable}

\begin{deluxetable}{ccc}
\footnotesize
\tablecaption{The eigenfrequency $\bar{\omega}$ 
for low-radial-order $g$-modes of the $n=1$ polytrope
with $\gamma=-10^{-4}$ at $\Omega=0$ \label{g-mode}}
\tablewidth{250pt}
\tablehead{
\colhead{$l$} & \colhead{$g_1$}  & \colhead{$g_2$}  
}
\startdata
 1 & $1.20\times 10^{-2}$ & $7.69\times 10^{-3}$  \nl
 2 & $1.68\times 10^{-2}$ & $1.15\times 10^{-2}$  \nl
 3 & $2.01\times 10^{-2}$ & $1.44\times 10^{-2}$  \nl
 4 & $2.26\times 10^{-2}$ & $1.68\times 10^{-2}$  \nl
 5 & $2.47\times 10^{-2}$ & $1.88\times 10^{-2}$  \nl
 6 & $2.65\times 10^{-2}$ & $2.05\times 10^{-2}$  \nl
\enddata
\end{deluxetable}

\clearpage

\begin{deluxetable}{cccccccccc}
\footnotesize
\tablecaption{Dissipative timescales $\tilde\tau$ of $r$-modes\tablenotemark{a} 
\  for the polytropic neutron star models with the index $n=1$. }
\tablewidth{0pt}
\tablehead{
\colhead{$m$} & \colhead{mode} & \colhead{$\gamma$} 
 & \colhead{$\kappa_0$}
 & \colhead{$\tilde \tau_B$(s)}     & \colhead{$\tilde \tau_S$(s)}     
 & \colhead{$\tilde \tau_{J,\vert m\vert}$(s)\tablenotemark{b}} 
 & \colhead{$\tilde \tau_{D,\vert m\vert+1}$(s)} 
 & \colhead{$\tilde \tau_{J,\vert m\vert+2}$(s)}
} 
\startdata
1&$r_1$&
$0$
 & 0.691 &$ 5.89\times 10^{9} $&$  9.23\times 10^{7} $
 &$ \cdots$&$ -2.46\times 10^{5} $&$ -1.27\times 10^{8} $
\nl
 & &
$-10^{-4}$
 & 1.000 &$ 6.08\times 10^{9} $&$  9.04\times 10^{7} $
 &$ \cdots$&$ -2.31\times 10^{5} $&$ -1.58\times 10^{8} $
\nl
 & & & & & & & & \nl
2&$r_0$&
$\phm{-}10^{-4}$
 &0.667 &$ 2.03\times 10^{11} $&$ 2.50\times 10^{8} $
 &$ -3.31\times 10^{0} $&$ -3.49\times 10^{2}$&$ \cdots$
\nl
 & &
$0$
 &0.667 &$ 2.03\times 10^{11} $&$ 2.50\times 10^{8} $
 &$ -3.31\times 10^{0} $&$ -3.49\times 10^{2}$&$ \cdots$
\nl
 & &
$-10^{-4}$
 &0.667 &$ 2.03\times 10^{11} $&$ 2.50\times 10^{8} $
 &$ -3.31\times 10^{0} $&$ -3.49\times 10^{2}$&$ \cdots$
\nl
 & &
$-10^{-2}$
 &0.667 &$ 2.10\times 10^{11} $&$ 2.54\times 10^{8} $
 &$ -3.31\times 10^{0} $&$ -3.49\times 10^{2}$&$ \cdots$
\nl
 & & & & & & & & \nl
 &$r_1$& 
$0$
 &0.517 &$ 6.47\times 10^{9}  $&$ 6.18\times 10^{7} $
 &$ -1.31\times 10^{8} $&$ -8.39\times 10^{4} $
 &$ -1.88\times 10^{6} $
\nl
 & &
$-10^{-4}$
 &0.667 &$ 6.79\times 10^{9} $&$ 6.01\times 10^{7} $
 &$\phm{-}2.18\times 10^{7}\ \tablenotemark{c} $
 &$ -7.05\times 10^{4} $&$ -2.06\times 10^{6} $
\nl
 & & & & & & & & \nl
3&$r_0$&
$\phm{-}10^{-4}$
 &0.500 &$ 6.64\times 10^{10} $&$ 1.43\times 10^{8} $
 &$ -3.17\times 10^{1} $&$ -1.88\times 10^{3}$&$\dots$
\nl
 & &
$0$
 &0.500 &$ 6.63\times 10^{10} $&$ 1.43\times 10^{8} $
 &$ -3.17\times 10^{1} $&$ -1.88\times 10^{3}$&$\dots$
\nl
 & &
$-10^{-4}$
 &0.500 &$ 6.64\times 10^{10} $&$ 1.43\times 10^{8} $
 &$ -3.17\times 10^{1} $&$ -1.88\times 10^{3}$&$\dots$
\nl
 & &
$-10^{-2}$
 &0.500 &$ 6.97\times 10^{10} $&$ 1.46\times 10^{8} $
 &$ -3.17\times 10^{1} $&$ -1.87\times 10^{3}$&$\dots$
\nl
 & & & & & & & & \nl
 &$r_1$& 
$0$
 &0.413 &$ 6.97\times 10^{9}  $&$ 4.78\times 10^{7} $
 &$ -1.86\times 10^{10} $&$ -5.30\times 10^{5}$
 &$ -4.07\times 10^{6} $
\nl
 & &
$-10^{-4}$
 &0.500 &$ 7.42\times 10^{9}  $&$ 4.63\times 10^{7} $
 &$\phm{-}2.39\times 10^{5}\ \tablenotemark{c} $
 &$ -4.37\times 10^{5}  $&$ -4.48\times 10^{6} $ 
\nl
\enddata
\tablenotetext{a}{We present dissipative timescales only for those 
that are unstable to gravitational radiation reaction.}
\tablenotetext{b}{We present dissipative timescales $\tilde{\tau}_l$ due to 
the gravitational radiation reaction only for
the dominant multipole moments.}
\tablenotetext{c}{For these dissipative timescales, we obtain positive
values, which should be negative by definition. The reason for the
inconsistent result may be because the extrapolation formula (\ref{tau2b}) 
for $\tau^{-1}$, which is obtained at $\bar\Omega=0$, does not necessarily work well
for arbitrary values of $\bar\Omega$. 
For instance, we find that
$\tilde{\tau}_{J,\vert m\vert}/\tilde{\tau}_{D,\vert m\vert+1} \sim 10^{3}$ 
for $m=2$ and 
$\tilde{\tau}_{J,\vert m\vert}/\tilde{\tau}_{D,\vert m\vert+1} \sim 10^{2}$ 
for $m=3$, where the dissipative timescales are estimated at 
$\bar{\Omega}=0.1$. 
This may suggest that 
the mass multipole radiation dominates the current one 
for the $r_1$-modes for large $\bar\Omega$. 
Thus, our conclusions on the stability of the $r_1$-modes 
extrapolated for large $\bar\Omega$
are not altered at all.}
\end{deluxetable}

\end{document}